# REVIEWING THE FUNDAMENTALS AND BEST PRACTICES TO CHARACTERIZE MICROPLASTICS USING STATE-OF-THE-ART QUANTUM-CASCADE LASER REFLECTANCE-ABSORBANCE SPECTROSCOPY

Adrián López-Rosales[1], Borja Ferreiro[1], Jose M. Andrade[1]*, Andreas Kerstan[2], Darren Robey[3], Soledad Muniategui[1]

[1]Grupo Química Analítica Aplicada (QANAP), Instituto Universitario de Medio Ambiente (IUMA), Universidade da Coruña, Campus da Zapateira, E-15071, A Coruña, Spain

[2]Agilent Technologies, Hewlett-Packard Str. 8, 76337, Waldbronn, Germany

[3]Agilent Technologies, 679 Springvale Rd, Mulgrave, VIC, 3170, Australia

*Corresponding author: fax: +34-981-167065; *andrade@udc.es*

**ABSTRACT**

Microplastic pollution studies depend on reliable identification of the suspicious particles. Out of the various analytical techniques available to characterize them, infrared transflectance using a tuneable mid-IR quantum cascade laser is a high-throughput state-of-the-art imaging option, specifically Agilent´s QCL-LDIR (Quantum Cascade Laser Direct Infrared imaging). Its conceptual grounds are reviewed, instrumental developments are discussed, along with a review of applications and best practices to overcome obstacles/difficulties in routine measurements, namely: the spectral range, the variation of some peak intensities with the particles size, effects of the size of the particles, processing speed, and avoiding the use of measurement aliquots. Objective procedures to avoid too many false positives when identifying spectra and to distinguish fibers and fragments are given. These practices open a path to QCL-LDIR measurement standardization and potential use for microplastics monitoring, as requested by many governmental bodies in charge of setting environmental protection rules.

## GENERAL INDEX

## LIST OF ACRONYMS

ATR: Attenuated total reflectance

DLaTGS: L-alanine-doped triglycine sulphate detector

ECHA: European chemicals agency

FPA: Focal-plane array detector

FTIR: Fourier-transform infrared spectroscopy

GC: Gas chromatography

HDPE: High density polyethylene

HQI: Hit quality index (sometimes termed match index or correlation index)

IR: Infrared

LDPE: Low density polyethylene

LOD: Limit of detection

μFTIR: Fourier-transform infrared microspectroscopy

MCT: Mercury-cadmium-telluride ($Hg_{1-x}Cd_xTe$) detector

MIR: Infrared spectroscopy in the medium (4000-600 $cm^{-1}$) region

MPs: Microplastics

PA: Polyamide

PC: Polycarbonate

PET: Poly(ethylene terephthalate)

PMMA: Poly(methyl methacrylate)

PP: Polypropylene

PS: Polystyrene

PVC: Poly(vinylchloride)

QCL: (Tuneable) Quantum-cascade laser

QCL-LDIR: Quantum-cascade laser-based infrared spectrophotometer (commercially, “laser direct infrared”)

## GLOSSARY

A glossary with a collection of some technical terms used throughout the paper can be found in the **Supplementary Material**.

# 1. INTRODUCTION

Microplastic particles (MPs) constitute a new type of pollutants present throughout almost all ecosystems, though they are especially prevalent in oceans and coastal areas. They are also becoming a real concern in agricultural soils [1–3]. The high consumption of plastic products in our society, especially single-use applications, and their inefficient recovery and scarce recycling led to a high occurrence of plastic debris worldwide [4]. MPs were typically defined as plastic fragments between 5 mm and 1 µm diameter [5], whereas the European Chemicals Agency (ECHA) established them as "solid polymer-containing particles, to which additives or other substances may have been added, and where ≥ 1% (w/w) of particles have (i) all dimensions 0.1 µm ≤ x ≤ 5 mm, or (ii) a length of 0.3 µm ≤ x ≤ 15 mm and length to diameter ratio >3" [6]. They are caused nowadays mainly by photodegradation and erosion of macroplastics (and are termed "secondary MPs" ), as their production for various industrial and cosmetic applications (known as "primary MPs") has reduced notably [7,8] thanks to legislative efforts in –e.g.- USA and EU [9,10]. Obviously, an accurate identification of plastic particles, both their quantities and their major polymeric constituents, is crucial to monitor them. This would assist in the understanding of important questions like their sources, their environmental fate and allocation, which polymers tend to (photo)degrade faster, etc. Despite MPs have been studied for some years there are too many open questions that hinder further regulation efforts, decision-making and, ultimately, regulatory efforts, and, while some progress has been made, similar comments made in 2018 [8] remain valid.

The most common techniques to characterize putative MPs include Fourier-transform infrared spectroscopy (FTIR) [11] using either attenuated total reflectance (ATR) [12–14], micro reflectance [15,16], or transmittance [17,18]; Raman spectroscopy [19,20] and pyrolysis-gas chromatography hyphenated with mass spectrometry [12,21,22]. All of them offer good results although at the expense of very long turnaround times and high laboratory workloads when facing big amounts of particles per sample, as in field samples. These limitations make some of them inadequate for routine monitoring purposes, where tens or hundreds of samples need to be considered. A critical literature review concluded that fast, automated techniques, based on imaging detection [23] and avoidance of manual work [24] are urgent for MPs analysis. Further, they need to be fast in order for laboratories to afford studing the very many samples that monitoring programs imply.

About 5 years ago, a novel IR instrumental analytical technique was commercialized, and it may represent a breakthrough in MPs characterization. Overall, since 2023, over 200 publications, have been released dealing with MPs analysis using LDIR, with an approximately exponential trend (see

**Figure SM1, Supplementary Material**). A comprehensive review of most of the QCL-LDIR applications is given in section 3.5.

Initially developed for pharmaceutical quality control of solid products, a tuneable mid-IR quantum cascade laser-based (QCL) infrared imaging system (commercialized as LDIR –Laser Direct Infrared- by Agilent) was "refurbished" for fast-imaging of reflective surfaces holding the sample particles, in a remarkable reduced timeframe. Other companies follow this way; e.g., Bruker launched Hyperion II, though with a different instrumental concept (more information can be found at [25]). Despite the recent QCL-LDIR commercialization an increasing number of articles employed it for MPs-related studies in samples from different ecosystems [26], including sediments [27], food items [28], agricultural soil [29], biological tissue [30] and drinking water [31]. Some practical considerations, including economical, revealed that QCL-LDIR compares favorably to other systems [32]; as a resume, QCL-LDIR costs like a top FPA-FTIR system and is cheaper than a Raman systems (with various lasers to take account of fluorescence effects).

This paper aims at introducing the conceptual bases of QCL-based infrared systems and, in particular, the LDIR device, as pertaining to the transflectance spectroscopy analytical techniques, and shows how some instrumental parameters may impact the results and affect the laboratory performance. This is relevant because, roughly, most publications focused only on the final measurement step. Evaluating the adequacy of an instrumental technique for a given purpose (here, characterising MPs) is far from trivial because the "success ratio" of many technologies and techniques available on the market has not been studied deeply. No analytical technique was validated thoroughly for MPs analysis because only a handful of interlaboratory exercises (or ring trials) took place. One of them was the EuroQCharm/Quasimeme/Norman Interlaboratory Study [33] on the Analysis of Microplastics in Environmental Matrices held on 2022. Up to 67 reports from a wide range of participants and instrumental techniques were submitted, including LDIR, µFTIR, Raman, IR-FPA, Pyrolysis-GC, Spero-QT (a QCL-IR microscope) and several of their variants. Unfortunately, not many conclusions could be drawn because of the very large variability of the results (not only between the techniques but within them) and because there was a factor of inherent variability since the samples were pretreated as per in-house methodologies, not unified previously. Somehow, that study agreed with serious lack of comparability reported after the first relevant interlaboratory exercise on MPs analysis [34], with almost the same range of analytical techniques. Interestingly, none of them was found to exceed systematically the others and all had some outstanding characteristic advantage. The authors attributed most of the problem to the individual interpretation of the preparation protocols and working methods. However, in our opinion two relevant issues were not considered in those ring trials: (a) setting a/some reliable match threshold/s to link an unknown spectrum to the spectral libraries and (b) the lack of a validation standard over all relevant techniques. This will be discussed in section 3.5.

Hence, there are relevant open questions in the scientific literature, like standardizing operational procedures or deciding on the acceptance of false detection ratios (which might be a function of the particle size), which particle size should be considered for routine monitoring programs, etc. About the former issue, ongoing efforts study the workflow of LDIR measurements and how to standardize it [35,36]; including oxidative sample treatment and LDIR characterization [37]. A reflection is in order here: is it possible to get "a general procedure" to determine MPs? The most extended opinion nowadays is that different procedures must be defined for different matrices and, even, for different categories of the same type of matrix. For instance, a procedure for fish would not be equally useful for mackerel than for salmon since the fat content is widely different and, so, the digestion of the organic matrix should be different [36]. So, the final MPs characterization is just a part of the complexity of the whole problem.

In the authors´ viewpoint, the study presented here is relevant because, as for any analytical technique, a correct setup of the QCL-LDIR instrument/workflow may largely influence the final identification and characterization of MPs, as suggested elsewhere [38]. Despite previous reports [35,37] studied scan speed, sensitivity, analysis time and spectral match index, some information will be presented here as well. Similarly, although the mathematical grounds of transflectance spectroscopy had been presented elsewhere [39,40] they did not consider some of the practical issues put forward here.

Thus, the present work discusses in a practical way important issues for the routine use of the reflectance-absorbance QCL-based LDIR methodology. The paper is organized in the following way: Section 2 presents the basic instrumental grounds of the QCL-based IR technique and poses some problems/difficulties and challenges related to it. Section 3 constitutes the main body of the paper as it discusses some experimental issues and suggests ways to address some of the problems outlined previously. In particular, it studies the relation between particle size and total absorbance. Then, we evaluated the extent of the penetration of the laser beam in the particles and its influence on the spectral intensities, which –to the best of our knowledge- is approached first time in the MPs literature. Some practical notes will highlight the importance of the instrumental sensitivity to detect particles correctly, the influence of the size of the scanning area in the total amount of detected particles, how to differentiate fibers from fragments, and how to set a reliable match index to identify unknown spectra (a novel tiered approach is proposed). This latter issue is of paramount importance to gain confidence on the reported values derived from QCL-LDIR measurements.

## 2. EXPERIMENTAL PART

## 2.1. Conceptual bases of a QCL-based LDIR instrument

### 2.1.1. The QCL source

The radiation source, the very core of any QCL-based instrument, is a tuneable QCL. Described for the first time in 1994 [41], it consists of a semiconductor laser emitting radiation in the mid-infrared region (MIR). It is smaller, offers a much higher energy output than any classical IR source, it is more easily tuneable than traditional lasers and can be operated at room temperature.

In a traditional semiconductor laser, a pure semiconductor material is excited either electrically or by an external source of radiation so that the electrons in the valence band migrate to the conduction band (these terms are defined in **Supplementary Material**). This is exemplified in **Figure 1a**. The energetic difference between the valence and conduction bands is called “gap” and plays a very relevant role when talking about lasers. After absorbing quantized energy from an external excitation source, an electron increases its energy state to a higher electronic, vibrational or rotational level. When that electron, in presence of a photon, is relaxed to a state of low energy another photon is emitted, with the same direction, and in **phase** with the first photon, whose energy (and, so, the wavelength) is determined by the semiconductor gap and the thickness of the semiconductor. Typical semiconductor materials are GaN, InGaAs/InAlAs, GaAs/AlGaAs, InGaAs/AlAsSb, InAs/AlSb, AlGaN/GaN, and AlN/GaN (using the names of the elements of which the materials are constituted) [42].

a)

b)

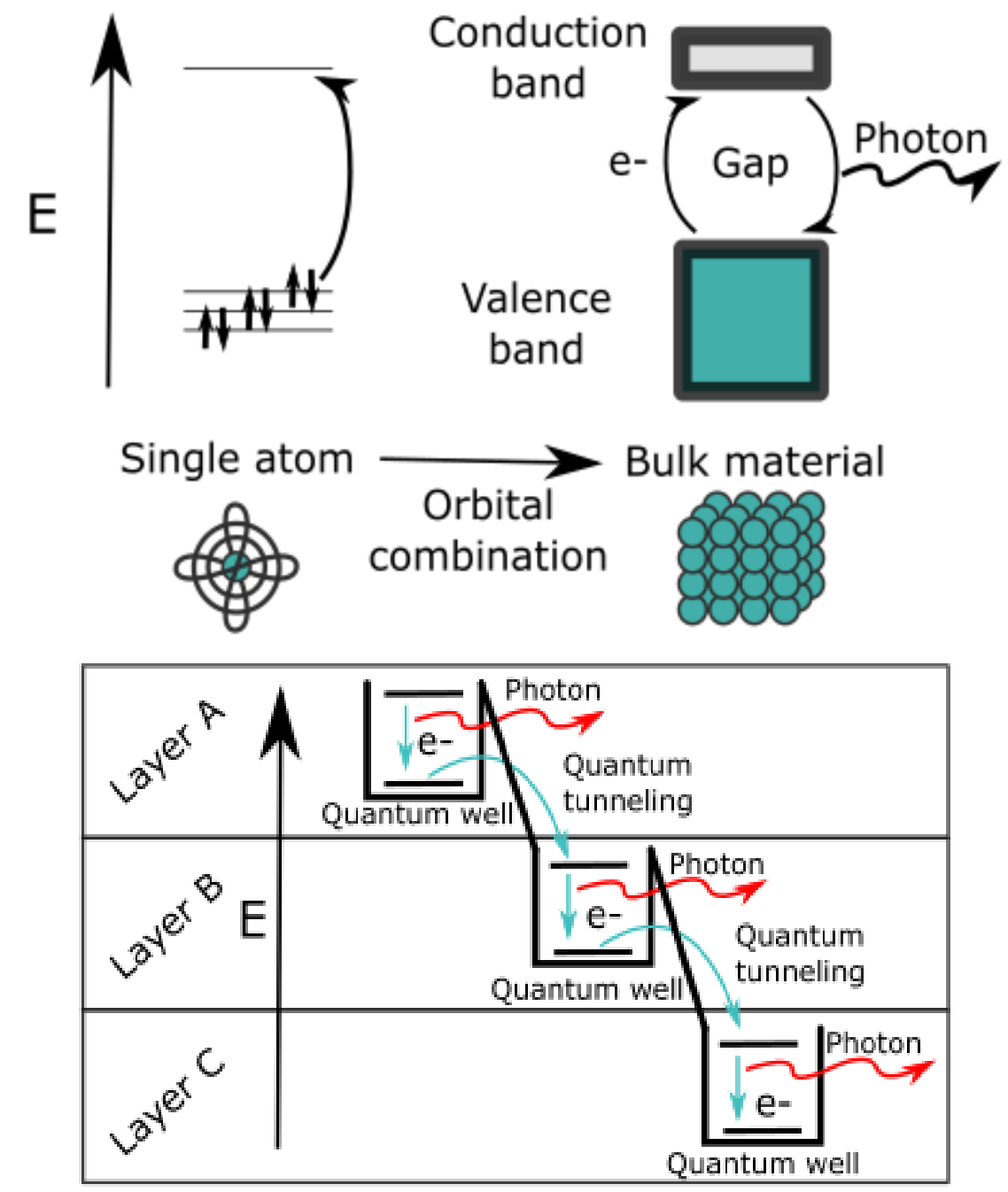

**Figure 1:** Basic concept of a QCL: a) Generic visualization of a semiconductor band model. The grey electrons ($e^-$) are shared with contiguous atoms of silicon in the lattice. b) Quantum cascade band model associated to a three thin-film-semiconductor-layer laser (E stands for energy level).

In a tuneable QCL however the laser is not composed of a single semiconductor but of several thin films of different semiconductor materials (or the same semiconductor material although with different thicknesses) ordered appropriately [43,44]. Their thicknesses are of the order of only some nm (the size of the De Broglie´s wavelength (see Supplementary Material) of an electron), to produce the so-called quantum wells or quantum confinement (see **Figure 1b**). These are very narrow spaces where the movement of the electrons and their associated electron holes (called excitons in semiconductors) is confined and, so, the energetic levels of the excitons became well-defined and restricted to discrete quantized values.

When the exciton relaxation is induced by a photon, the newly produced photon adopts the direction of the original one, generating a highly focused, polarized beam. Radiation released from layers of different semiconductors, or different thicknesses of the same one, yield different wavelengths (so, the term “tuneable”). When several quantum wells are placed close each other a succession of quantum wells is created. In each well an electron (or an exciton) can relax from a high energy state to a fundamental one and then move to the next quantum well, whose highest energy state is close to the lowest of the preceding well (this effect is called quantum tunnelling). There, the electron relaxes again from the highest to the lowest state, and so forth, like in a cascade, emitting a photon in each of these “jumps” [45] (**Figure 1b)**, thus generating a beam of collimated, polychromatic high intensity light. Interested readers can consult more details in some general-purpose overviews on QCL [42]. For the QCL-based IR system we are dealing with in this tutorial, the laser is of a constant wave type although operated in a pulsed mode (i.e., offering discrete bursts of light spaced on time at a programmed repetition rate).

The differential characteristics of a QCL-based system can be summarized thinking on the laser properties. A traditional FTIR thermal multiwavelength source emits photons over a relatively broad spectral range; however, the number of photons per wavelength (spectral power density, see **Supplementary Material**) is quite small. A QCL on the contrary, for a given setup, emits all its photons on approximately the same wavelength [46], as a typical laser. This means that the spectral power density of a QCL is typically orders of magnitude higher than that of a thermal source [46] and even of a synchrotron [47].

Just for comparison, it has been reported that QCLs outperform diode lasers (operating at the same wavelength) by a factor >1000 in terms of power [42], and Globar sources (by a factor of 10,000 in terms of spectral emission power [48]). However, one must be cautious with some comparisons reported in literature because some terms can be mixed or misleading. For instance, brightness is often confused with "power" ; brilliance is confused with spectral intensity. Spectral power density (or spectral power distribution) is mixed with spectral power distribution. See **Supplementary material** for a definition of all these terms.

Another relevant advantage of QCLs is the remarkable light coherence they offer, allowing for particularly small linewidths (typically <0.001 $cm^{-1}$) compared to other mid-IR sources [49] and, so, high focusing capabilities, around 5.5-12.5 $\mu m^2$ for a spot (1800-975 $cm^{-1}$ spectral range) *vs.* 20-50 $\mu m^2$ of classical IR sources (ca. 5000-400 $cm^{-1}$ spectral range). Such focusing capabilities allows QCL- LDIR systems to –in principle- measure down to 4-5 µm particles. However, for current uses it is recommended to limit the automatic detection of particles to 10 or 20 µm [35] because, at the instrumental limit of detection, the smaller particles might not be located. This is a problem common for all technologies, and it is the concept of the limit of detection, but very few studies have been done in relation to MPs. In addition, in many field samples the number of particles increases exponentially with decreasing size [50] and, therefore, the measuring times would be unpractical (see section 3.5) if the very small particles (<20 µm) are to be characterised. A consequence of the high focusing capability of QCL is that it enables higher penetrations (path lengths) than FTIR sources.

### 2.1.2. From polychromatic to monochromatic radiation

In order to get a spectrum, the sample specimen has to be irradiated sequentially with a collection of sequential linewidths (each comprising a as-small-as-possible range of wavelengths around the nominal value) spanning through the spectral range available for the instrument. This means that the QCL beam has to be decomposed into its energy components. A possibility to get a small linewidth radiation is to modify the frequency range of the waves emitted by the laser by adjusting the power applied to the QCL, although this has several drawbacks [51]. Another more efficient possibility, applied in the present system, is to generate multiwavelength radiation with the QCL and tune it afterwards. This can be made in different ways, a common one couples the QCL source itself with an external cavity (this option provides a broad tuning range, up to several hundred wavenumbers) [48] which includes a dispersion device. That can be of different types: a Fabry-Perót system [48] (or Etalon device [51]), a diffraction grating in a Littrow configuration [52], a Bragg grating [49], a rotating diffraction grating [46], a collection of filters, etc., along with dedicated optics to focus the light [53]. For the purposes of this tutorial, all these subsystems (radiation decomposition device plus optics) are incorporated in a module labelled as "QCL" (following what has been called the QCL chip [49]) in **Figure 2**. There a conceptual design of a QCL-LDIR instrument is presented, depicting its most relevant modules. It can perform both transflectance and traditional ATR measurements (a major application when it was first developed for the pharmaceutical field, as mentioned above) thanks to a set of movable mirrors that redirect the light to the proper optical path.

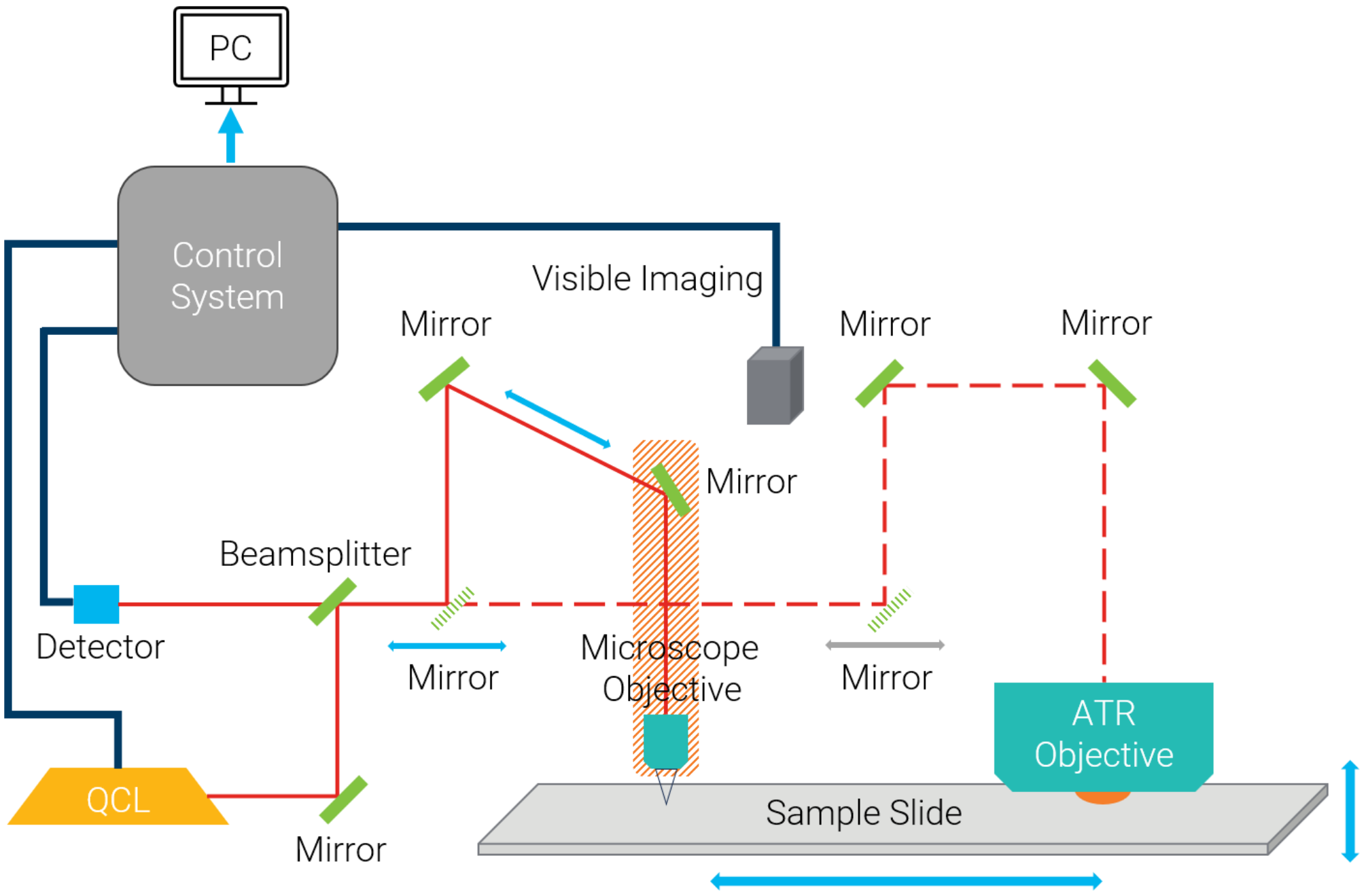

**Figure 2**: Conceptual scheme of the 8700 QCL-LDIR from Agilent. © Agilent Technologies, Inc. Reproduced with Permission, Courtesy of Agilent Technologies, Inc.

### 2.1.3. Detection

After irradiating the particles, the laser beam is directed back (**Figure 2**) to a thermoelectrically refrigerated mercury cadmium telluride (MCT, $Hg_{1-x}Cd_xTe$) detector (also a semiconductor thin film), protected with an IR transparent window. MCT detectors are sensitive and high-throughput devices that allow for fast measurements. Their working principle is that when a photon with a sufficient energy impacts the semiconductor, an electron of the valence band ascends to the conduction band [54]. After applying a voltage, the electrons in the conduction band generate an electrical current that is proportional to the intensity of the received infrared radiation. This can be related directly to the radiation absorbed by the sample (as defined by the Lambert-Beer-Bouguer absorbance equation: $A = -\log_{10}(I/I_0)$, see **Suplementary Material**).

This type of detectors can be active in a 4000 – 400 $cm^{-1}$ range, though this depends on the specific composition of the semiconductor [54]. In the LDIR instruments the range is reduced to 1800-975 $cm^{-1}$ to partially encompass most of the fingerprint region of the IR spectrum (1400-600 $cm^{-1}$), unique for each polymer, and to increase the measurement speed per particle. A drawback of semiconductor detectors is that if the infrared radiation is too intense all available electrons are promoted to the conduction band and the detector will be saturated, not being able to sense any further increase in intensity. To avoid this in LDIR systems light polarizers are employed [55], that reduce the IR energy throughput significantly [56], hence, they are low-throughput radiant power systems (i.e., less than 20 % of the IR beam reaches the detector). Further, they are fast-scanning systems that spend around 1 s in getting a spectrum on each particle [57]. As a matter of comparison with FTIR, the response of the common DLaTGS (L-alanine-doped triglycine sulphate) detector for the medium-IR region is reduced by half for every two-fold multiple of the scanning velocity [58]), although this does not happen with MCT detectors. A straightforward comparison of different common IR detectors and some of their capabilities can be seen elsewhere [58]. In addition, the QCL-LDIR systems feature a distinctive characteristic when compared to FTIR ones, mostly with those based on FPA (focal plane array) detectors, which are used frequently to characterize MPs. For an FPA to be effective, all its pixels need to be illuminated equivalently. This implies defocusing the IR beam and getting a general image although without, strictly speaking, focusing on the measured particle. On the contrary, the QCL beam is focused on each particle being measured.

High intensity laser sources can contribute to the overall measurement noise (particularly when operated in pulsed mode due to pulse-to-pulse intensity fluctuations) and this effect partially cancels out the relevant signal-enhancement due to their energy [48]. This is a drawback shared by all high-throughput IR measurement devices (FPAs, IR microspectroscopy, etc.) as reducing the number of scans is an effective way to speed up measurements. The QCL-LDIR system takes account of this problem by using a second detector (not shown in **Figure 2**, in a perpendicular position to the detector shown there, after the beamsplitter) and avoiding data collection between the light pulses.

Finally, when transflectance measurements of thin films are made at near normal incidence angles (like LDIR does) the spectrum is dominated by absorption processes since specular reflectance from the outer surface results in only 4-10 %. For this reason, transflectance spectra of very thin films resemble very well transmission/absorption spectra, as far as the sample thicknesses are in the 0.5 - 20 μm range [59].

### 2.2. Sample-radiation interaction: transflectance

To measure a specimen with the LDIR it has to be located over an inert, totally reflecting, flat surface. When collimated IR light irradiates the surface of the specimen it can be reflected without entering the sample or can be reflected after penetrating it. These two reflection modes are termed surface-and volume-reflection, respectively [60]. In general, both occur at the same time, affecting the spectral features with proportions that depend (at the very least) on the thickness of the sample, the surface conditions (roughness), the angle of incidence of the radiation, its polarization, structural composition of the radiation and the optical properties of the material under study. **Figure 3** summarizes these phenomena. In surface reflection, the IR beam can be reflected in (almost) a purely specular way (specular reflection or just, reflectance; whenever the incidence area is flat and smooth) or in a dispersive form (diffuse reflectance, when rough, uneven surfaces or powders are measured). When light penetrates the sample (volume reflection) it can be absorbed, refracted, reflected (at the reflective supporting medium) and scattered prior reaching again the surface and coming out as an overall reflected beam.

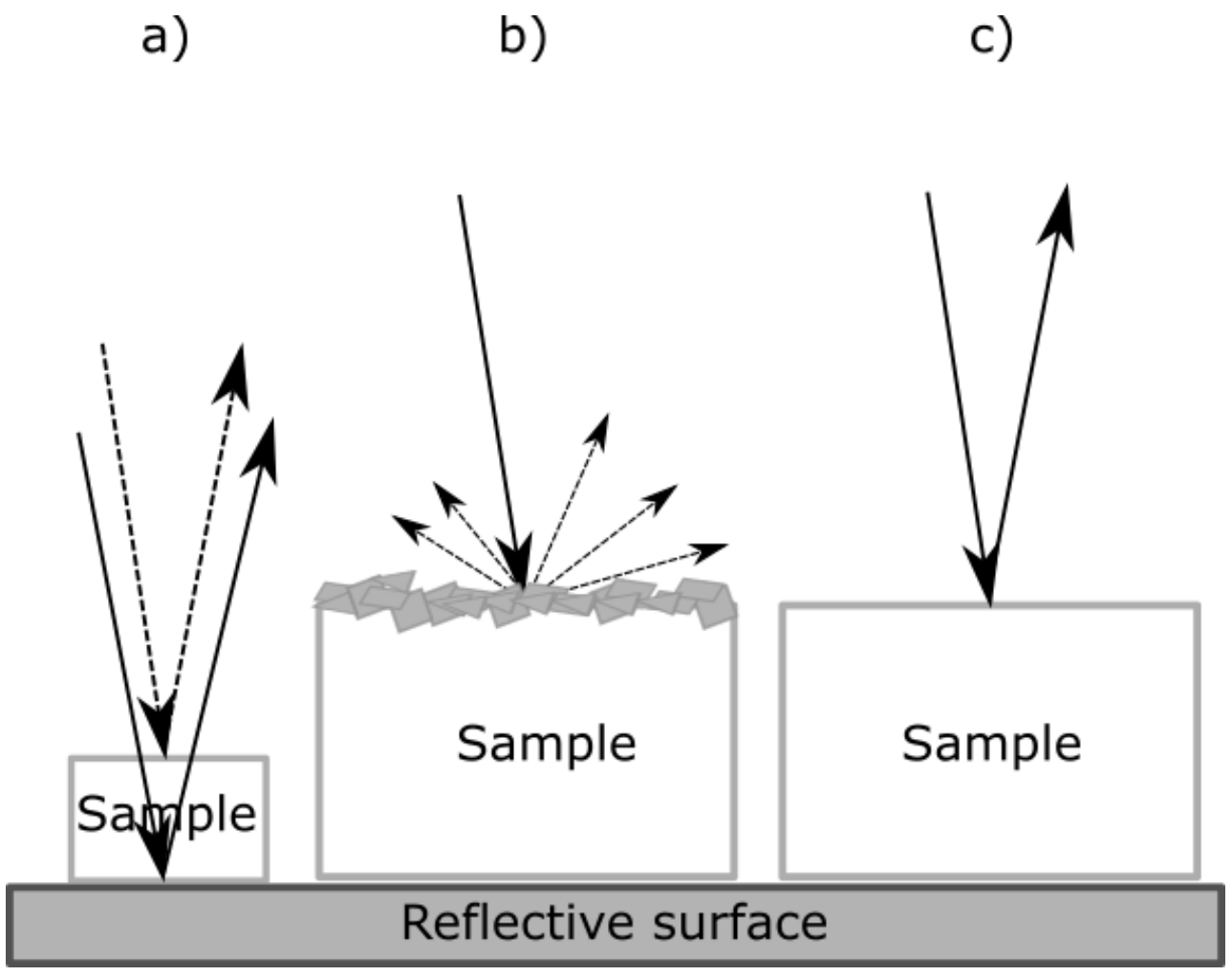

**Figure 3**: Exemplification of transflectance (a), diffuse reflectance (b) and specular reflectance (c) processes that can occur when a particle over a highly reflective surface is irradiated with a quantum cascade IR laser.

The analytical technique that uses the penetration of the light and its final emergence as a reflected beam receives several names, of which transflectance, as it was defined first time [61] is used most commonly in the MPs field (see **Suplementary Material** for more information). Recall that reflectance signals do not hold the linear behaviour of the Lambert-Beer-Bouguer's law (as it happens in other types of spectroscopy), but the absorbance ones do and, so, it is good practice to transform reflectance to absorbance-like spectra. A relevant warning is in order here: Transflectance is neither specular reflection (because in that case their appearance should present first-derivative shapes and, on the contrary, the experimental spectra resemble largely the transmittance/absorbance ones) nor a “simple” double transmittance [61]. In particular, the former consideration applies to thin films. Some specific detectors might cancel out the advantages of QCL-based sources and yield transflectance spectra quite different from their FTIR counterparts [62–64], see **Supplementary Material** for more details.

It is worth noting that despite there is a large body of knowledge for transflectance carried out on thin and very thin films (several nanometers thick) that is not the case for transflectance made on large items/layers [65]. Therefore, despite all efforts made by the authors, only a handful of references were found studying the theoretical and practical aspects of some issues discussed here on “not-thin” particles, as MPs can be considered. Although a mathematical treatment of transflectance is far beyond the scope of this paper, some remarks are needed as they are required for subsequent explanations. Interested readers are kindly forwarded to the references included in this section.

In classical spectroscopy the absorption bands (which are related logarithmically to the transmittance) are due to the absorption (or extinction) coefficient (*k*). However, this is only true for situations where no scattering at all occurs (e.g., transparent and low concentration solutions). When scattering occurs, absorptance (α) is a more appropriate term, which is the ratio of the absorbed to the incident radiant power; when $\alpha \leq 1$, $\alpha \approx A_e$, being $A_e$ the Napierian absorbance [66]. For decadic absorbance this may be symbolized as: $A_{10} = -\log_{10}(1-\alpha)$. When a sample transmits and reflects light the fractions of light absorbed (α), reflected (R) and transmitted (T) amount 1 (or 100 % incident power), so that $\alpha + R + T = 1$; this is the energy conservation law or Kirchhoff”s law of radiation.

The index of refraction, *n*, defined as the ratio of the speed of light in vacuum to that in a given material, is far from simple and does not show a linear behaviour. Indeed, the complex index of refraction (*n*’) is defined as $n' = n - i \cdot k$. Where i is the complex number $\sqrt{-1}$ and, so, the term ($-i \cdot k$) is called the imaginary absorptive part of the complex refractive index. In IR spectroscopy this part is significant and therefore the real part of the complex index undergoes a large anomalous dispersion (with consequences on the spectra). It is this complex interplay of absorption, anomalous dispersion and energy transport that, in part, complicates the understanding of the recorded data [39,40]. Recall that these phenomena are not independent, all relate to the same overall physical process in the sample material and occur at the same time (radiation-sample interaction).The chemical composition of the sample is the critical factor for its optical properties in terms of absorption and refractive indices, influencing the degree of penetration of the radiation and, so, the contribution of the different radiation interactions.

Most micro-spectrophotometric infrared devices, including the LDIR system, can be used in reflectance, reflectance-absorbance (transflectance) and attenuated total reflectance (ATR) modes, the latter using a highly refractive Ge crystal (see, e.g. **Figure 2**). However, the majority of MPs-related studies will be done in transflectance mode –as far as the particles are partially transparent to the corresponding IR wavelength- and, so, it will be the only mode considered in this work. A major reason is that small particles might stick to the ATR tip and make practical measurements unfeasible due to the need to clean the contact tip after measuring each particle. Hence, a pragmatic proposal might be to reduce the overall cost of the LDIR system and its optical complexity by eliminating the ATR part when dedicated to MPs studies.

As a rule of thumb, a “typical” transflectance process will dominate when small particles are irradiated thanks to the highly focused, radiant intense laser beam, though in big or opaque particles external reflectance and/or volume-absorbance might be prevalent. Right now, there is not a given threshold to define “big” or “small”, although $>\lambda/4$ is a common threshold in physics to differentiate “big” and “large”. Thus, if two exemplary wavenumbers of the working range of a LDIR system are considered, 1000 and 1800 $cm^{-1}$, they correspond to 10 and 5.5 µm, respectively. Therefore, it seems clear that the particles we usually measure using a LDIR should be considered “big” from a radiation viewpoint.

This issue is relevant because although not considered so far it does impact the final spectra of putative MPs; see section 3.2. If transflectance occurs predominantly a typical transmittance-like spectrum is expected whereas if reflectance predominates other phenomena may occur and they, sometimes, are visualized as bumps and broad spectral bands. Besides, superficial external specular and diffuse reflectance depend on the morphological characteristics of each particle.

In classical IR systems the recorded spectra are different for each mode and specific spectral treatments (Kubelka-Munk and Kramers-Kronig) are required once the spectra were obtained (a task to be done on each spectrum by the analyst) to obtain absorbance-like spectra. The LDIR system includes the Kramers-Kronig calculation which assumes a –essentially predominant- specular reflectance. However, it is difficult to evaluate whether the specular part of the reflection is prevalent. A purely pragmatic way to get an idea on how high the portion of reflectance is for a spectrum consists of using the Kramers Kronig (KK) transformation available in the software. If the KK spectrum and absorbance spectrum look similar, the particle is more transparent and in both cases the spectrum appears like a transmission spectrum. Otherwise, a large proportion of external (diffuse) reflectance may be suspected. Probably, for roughest particles external diffuse reflectance will be important. Therefore, one should be aware of this problem when studying the match of unknown particles to a spectral library. An instrumental strategy used in LDIR systems to mitigate this problem is depicted in the **Supplementary Material**. It consists of a built-in device that separates the specular and the diffuse reflectances by performing polarization measurements at a point of interest [55,56,67].

### 2.3. Samples and sample deployment

The polymers used throughout this study represent 9 of the most ubiquitous plastic types, accounting for >74 % of the plastic demand worldwide [68]. They appear very frequently as residues, and their presence is evaluated in all environmental studies: high density polyethylene (HDPE), low density polyethylene (LDPE), polyamide (PA), polycarbonate (PC), poly(ethylene terephthalate) (PET), poly(methyl methacrylate) (PMMA), polypropylene (PP), polystyrene (PS), and poly(vinylchloride) (PVC). They were provided by the Universität of Bayreuth (Germany) in the framework of a Joint Initiative Program Oceans-EU-funded project, BASEMAN [69]. The polymers were obtained as pellets (around 4-5 mm diameter) and powders (around 300 µm, but for PVC, which was around 80 µm size) and contained only the minimum number and quantities of additives required to get the polymer (i.e., they had no radical scavengers, colours, fillers, plasticizers, etc.), as requested in the project. They were fabricated by INEOS Styrolution (PS, commercial name: Styrolution PS 158 N/L); Borealis (PP, commercial name: HL508FB); Vinnolit Gmbh (PVC, commercial name: Vinnolit S3268); Neogroup (PET, commercial name: Neopet 80); LyondellBasell (PE, low density polyethylene, commercial name: Lupolen 1800P); and BASF (PA6.6, commercial name: "Ultramid"). In order to expand the range of particle sizes employed in this study those original forms were grinded down to various sizes in our laboratory, reaching a 2 mm to 20 µm size interval, using silicon carbide sandpaper.

All the assays presented here were made with the 1.6 version of Clarity Software in an HP Z4 G4 Workstation® with an Intel Xenon 3,6 GHz (4 Core 8 treads) processor, RAM Memory 64GB DDR4, GPU AMD Firepro W2100 2 GB and High Disk 500GB SATA3. FTIR studies were done with a Spotlight 200i Perkin Elmer microscope (micro reflectance measurements), using a Globar source, coupled to a Spectrum 400 Perkin Elmer FTIR instrument (microscope aperture = 50x50 µm – adjusted when required-, 50 scans/sample, Strong Beer-Norton apodization, nominal resolution 4 $cm^{-1}$, background correction before each measurement, 4000-600 $cm^{-1}$ range, Kubelka-Munk transformation). The microscopy has an MCT detector, cooled with liquid nitrogen.

Two options are available to place the particles in a LDIR system: (i) rectangular reflective microscopy-like slides (MirrIR slides, from Kevley Technologies), with a silver coating which allows an almost complete IR light reflection on the 400 - 4000 $cm^{-1}$ range [70–72]; and (ii) PC gold-coated filters (commercialized by Sterlitech and SPI supplies, US), and PET gold-coated filters: i3 TrackPor P (commercialized by i3 Membrane, Germany), which can be placed easily in dedicated holders [73][74]. Aluminium is a very interesting reflecting surface as suggested elsewhere [75] and filters of this material are being released also by i3 Membrane for some specific applications and they may represent a cost-effective alternative.

It is worth noting that when field samples are considered different sample treatments involving flotation, filtration, matrix digestion, etc. [29,76,77] are required to get rid of the interfering matrix and release the MPs. In that sense, Lusher et al. [78] , Prata et al. [79], Shumway et al. [80], and Bellasi [81] reviewed different protocols to tackle sample treatment issues, while Enders et al. [82] established a general guideline to approach complex samples. Buy they did not considered LDIR nor the need to transfer the sample to reflective surfaces.

This step is very complex and out of the scope of the present work so it will not be considered in depth. But the reader must be aware of this point and the fact that the final aliquot to be measured will contain organic and inorganic remains/particles in addition to the MPs under investigation. Hence, placing the particles of interest over the reflective surface can be challenging.

Previous applications reported resuspensions of the filtrate from the sample treatment in a 50 % water:ethanol mixture and withdrawing an aliquot of the total volume to the slide [77,83–85]. However, this can affect representativeness, mainly if very little volumes are considered [36]. Another possibility is to pick up the particles from the filter to the Kevley slide with the help of a stereomicroscope and micro tweezers. More recently, a third alternative to avoid withdrawal of aliquots was proposed based on evaporating the re-suspended filtrate solution with a Syncore evaporation system and placing the whole concentrate over the reflective surface. The three approaches were compared, and it was seen that “aliquoting” yielded poor reproducibility, and that the manual pickup is only suitable for non-fibrous particles over 70 µm (which, nevertheless, is a reasonable limit for many monitoring studies as current oceanic monitoring uses manta trawls of about 300 µm mesh size). The best alternative was the “Syncore method”, with very good recoveries for particles and fibers (around 90 % and 76 %, respectively), very reproducible, capable of processing up to 12 samples at a time without cross contamination; however, it is a slow process, and some days are required [36]. This was the approach followed in this work whenever Kevley slides are referred to. More specific details can be consulted in the references cited above.

The arrangement of the particles in the slide is relevant. When samples contain many particles (as it has been the case in our experience) it is very convenient to spread them throughout the whole reflective surface, not accumulating the drops at just one location. The latter option might be useful if few particles are expected. Kevley slides can also be affected by remains of the oxidative treatments as the silver coating degrades. This was clearly photographed recently [37] and we experienced the same problem. Kevley Tecnologies kindly informed us that degradation leads to, most likely, silver oxide particles (they also might be silver chloride if some form of chloride is in the oxidant solution). When the slides start to degrade, we noticed that the LDIR system detected too many tiny particles (that at the end were unidentified) and they may be due to those degradation particles.

Some authors tried to avoid aliquots evaporating the suspensions under a $N_2$ flow [30,86–90], or using special ovens [91]. Other options were not explained [88,92–94].

The use of metal-coated filters (gold or aluminium) avoids the need for evaporation processes or drawing aliquots. This may save time and simplify the working protocols as far as the filtrate is not too crowded with particles [35]. As a consequence, this requires a careful selection of the initial quantity of sample and scaling the working protocols accordingly [74]. There might be some concern about the high price of gold-coated filters, but be aware that over the last two years the prices dropped down by 50 % with a strong positive impact on the consumable cost per analysis.

In any case, avoid overlaying particles when depositing them in the slide/filter. Avoid too crowded filters by filtering through more than one, clogging is usually detected because of too lengthy filtrations. In case of problematic samples use stainless steel filters, mesh size 20-10 µm, to simplify final filtering [95,96]. Another option whenyou are aware of the existence of big and small particles in the same sample is to fractionate the particles by size before depositing them in order to have the best possible focus for both. A general scheme of the sample treatment procedures considered in the works developed by the authors (and referred to in this paper) can be seen in **Figure SM2(a)**, **Supplementary Material**.These protocols address one of the issues mentioned in Annex F of the ISO/DIS 16094-2 draft regarding the difficulty and risk of particle loss during transfer processes to reflective slides [97].

### 2.4. Good measurement practices

At the beginning of the LDIR measurement workflow a lateral topographical profile of the holder plus the particles is recorded. The maximum height of the particles deposited over it helps focusing the camera. Then, an image of the plate in the visible region reveals the general distribution of the particles, which may suggest the analyst to focus on a given region of the plate and save time as other parts may not be measured. Such an image is possible because the reflective surface and the particles reflect/absorb light differently (in both the visible and IR region). The most updated control software allows the analyst to carry out an initial screening irradiating the plate with an IR monochromatic light (e.g., 1442 $cm^{-1}$ to visualize the typical C-H bending of the polymers, or any other wavenumber of interest). This is very interesting to discriminate potential MPs from other particles and results in a notion of their arrangement in the plate. It may even be possible to set the analysis only on the particles that yield a signal at a given wavenumber; this strategy is termed “peak analysis” or “single absorption peak analysis” [62], depending on the company. As for every mode of IR spectroscopy, a background is required to get the net spectra of the particles. This is obtained by measuring the spectrum at an empty position over the reflecting surface nearby the detected particle(s). As the measurement of the slide progresses, several backgrounds are recorded and used subsequently.

Contrary to most FTIR measurements, where a large number of scans is averaged to improve the signal-to-noise ratio, the high radiant energy (brilliance) of the QCL beam allows spectra to be collected routinely with just one scan, with less noise than its FTIR-counterparts (hence, the possibility of reducing the measuring time). The system records a “full” spectrum (1800-975 $cm^{-1}$) in 1 s and requires ca. 4 s to process and match it and relocate the optics. Besides the highly-focused monochromatic laser beam, the detector has also a role here as the MCT ones are especially suitable for high-throughput measurements, as discussed in section 2.1.3. Therefore, the analyst has to be aware that noise averages are not done unless requested.

Finally, after a spectrum is measured it is compared with a previously selected spectral library. An identification is suggested, along with its corresponding correlation/match index (called here HQI, hit quality index). It is highly advised that the analyst incorporates additional spectra to the databases, see section 3.5. Ongoing advances in software would improve this issue.

Fibers, a quite difficult kind of MPs, can be characterized best by analysing several points along them (“line profiling”). As this is a particle-by-particle work that demands time from the analyst, it is only recommended in specific points or to get more confidence in particular identifications.

## 3. RESULTS AND DISCUSSION

In this section several major questions related to the LDIR workflow and how to obtain the best possible outcomes are studied:

(i); Is the spectral working range enough to identify polymers?; (ii) does particle size influence the transflectance LDIR spectra?; (iii) to which extent the QCL radiation penetrates into the MPs?; (iv) does the sensitivity parameter affect the overall detection capabilities?, and is it possible to improve instrumental resolution and visualize small particles?; (v) is it possible to reduce the turnaround time when many particles are to be investigated?; (vi) to what extent a spectral match threshold (HQI) can affect MPs identifications?; (vii) is it possible to differentiate fibres and fragments with the geometric parameters offered by the software at present?

### 3.1. LDIR spectral range

A common concern about QCL-LDIR systems is the limited spectral working range (1800-975 $cm^{-1}$), mostly when compared to MIR and/or NIR spectroscopy. However, it encompasses most of the "fingerprint region" although the lower limit of 975 $cm^{-1}$ excludes interesting spectral features like the out-of-plane C-H bending, and some other interesting vibrations.

A recent report considered this issue and found that the identification of most common polymers is not affected by the reduced spectral range [98]. To evaluate whether this might be a critical issue for polymer identification an indirect study was undergone. Multivariate statistical techniques (like principal components analysis, classification and regression trees, support vectors machine, etc.) were applied to search for the most relevant spectral regions required in traditional FTIR measurements (ATR and reflectance) in order to differentiate among common polymers [99]. It was found that most statistical models chose many wavenumbers in the fingerprint region affordable by the QCL-LDIR system, see **Figure 4**. The region between ca. 1800-2750 $cm^{-1}$ presents no absorption bands for usual polymers.

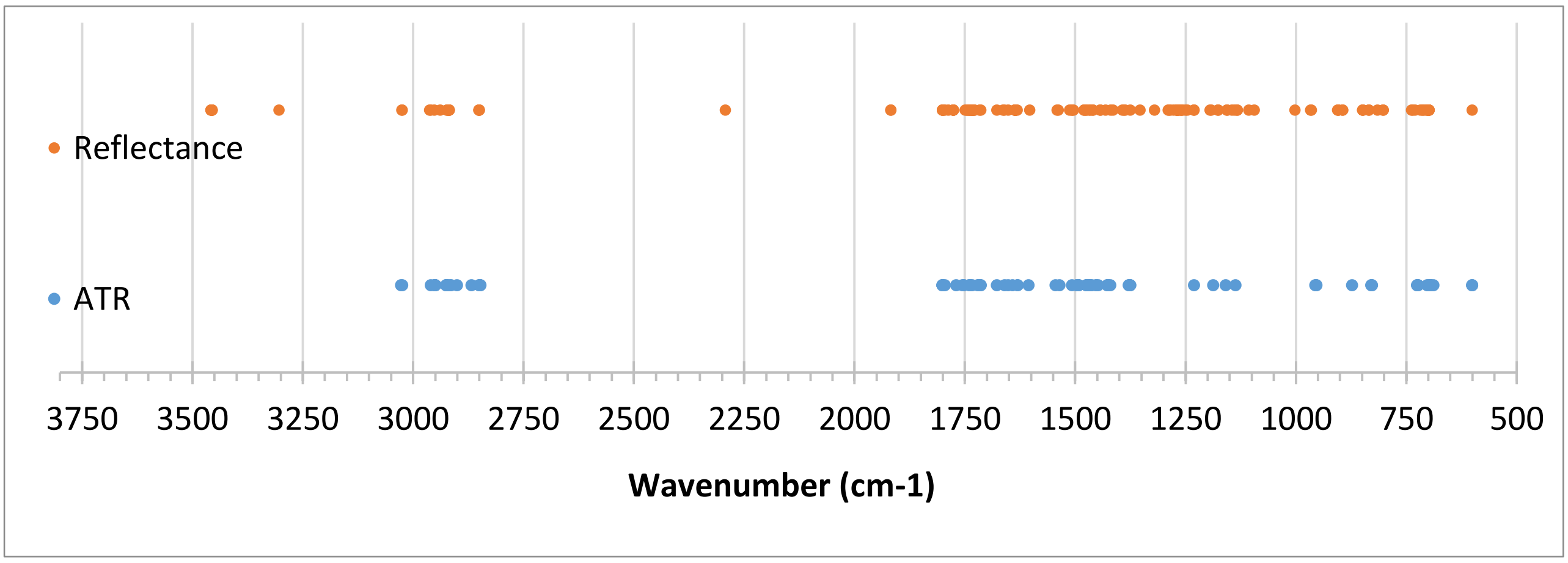

**Figure 4**: Graphical summary of the most relevant wavenumbers selected by several multivariate statistical models to differentiate nine common polymers using FTIR spectroscopy(more details can be found at [99]). Two measurement modes were considered. The spectra were obtained with a Spotlight 200i Perkin Elmer microscope (reflectance measurements) coupled to a Spectrum 400 Perkin Elmer FTIR instrument; the ATR measurements were made with a MIRacle one-bounce horizontal diamond ATR accessory (plus a Spectrum 400 Perkin Elmer FTIR instrument).

Similar conclusions were unveiled by other authors. For instance, a pulsed mid-IR QCL system operating in a very reduced range (1789.87-1350.07 $cm^{-1}$), working with a macro-ATR accessory, combined with linear discriminant analysis classified correctly (97 % success) five plastic types (PET, HDPE, LDPE, PP and PS) [100]. In the biochemical field good quantifications were achieved combining a QCL restricted to the 1200-950 $cm^{-1}$ range and partial least squares multivariate regression [48]. The region is also capable of differentiating natural from synthetic PA, which is a rather tough problem [101]. Interestingly, other suppliers of analytical instrumentation launched a new system with a reduced spectral range as well [25].

As a consequence, the spectral information derived from QCL-LDIR measurements seems sufficient to differentiate among the most frequent polymers. Likely, chemometric tools will be required to get best results in some cases. The accuracy of the correct identification in field samples will also increase whenever a suitable spectral database is employed. This issue becomes critical whenever multivariate statistical classifications or machine learning tools are used (some recent applications can be seen [102,103]) and it will be considered in section 3.5.

### 3.2. Particle size influences spectral intensity

No information was found in literature on the penetration of the radiation (nor path lengths) for QCL-LDIR or reflectance in FTIR systems in MPs. In general, the IR penetration depth is determined by the wavelength and the refractive index of the material. However, this might not be the only factor when the path length is considered. As an example, FPA detectors (current state-of-the-art in FTIR), need all their pixels to be illuminated homogeneously to work properly. So the IR beam needs to be defocused (as mentioned in section 2.4) and this spreads the IR energy over the whole FPA area. Although the total IR energy rendered by the source remains the same, each pixel receives less (than when no defocusing is used), requiring longer measurement times and lower S/N ratios. With QCL an approx. 5 $\mu m^2$ highly focused, monochromatic, coherent laser beam with a large spectral power density incides on the particle and may reach longer penetrations than a defocused FTIR beam. This, however, still needs to be demonstrated in practical terms.

Several reports from other fields than MPs inform on this issue albeit in different experimental setups. In many cases the path length depends on the configuration of the instrument or the sensor device and generalizations are not possible. Reported QCL path lengths for ATR were similar to those obtained with traditional FTIR devices: from 0.25 to 0.80 µm at 2500 and 700 $cm^{-1}$, respectively, using Ge windows [104]; and around 4-5.4 µm, 1200-900 $cm^{-1}$ range, when using a ZnS crystal to measure glucose and albumin in aqueous solutions [48]. This is because depth penetration in ATR is due to the evanescent wave. When other measurement modes were considered, larger path lengths and robust transmission measurements were reported, even in the presence of water (a high IR-absorber) [49], down to 165 µm [105]. Depth penetrations larger than conventional ones (around 38 µm; approximately 4x conventional FTIR systems) were observed when using external-cavity QCL measurements to determine proteins in aqueous solutions at the amide I band [47]. Finally, an outstanding depth penetration of 500-800 µm in aqueous solutions (with signal-to-noise ratios of 4:1) was achieved using a tuneable QCL laser built upon an amplifier [106].

Nevertheless, MPs measurements are done without solvents so, what happens when a QCL beam is focused on solid materials instead of solutions? We only found information in Medicine, where QCL is applied directly over the skin to seek for underneath analytes. Of course, the strong IR absorbance of fat and water represents a challenge there. Schwaighofer et al. [49] reviewed the measurement of bodily fluids and tissues to quantify selected analytes and reported path lengths around 100 – 165 µm in the 950-1200 $cm^{-1}$ region. In a different application, a QCL operating at 1750 $cm^{-1}$ ablated biological soft tissue and reached penetrations of 29 and 66 µm for wet and dry tissue, respectively [107].

As indicated above, information about path lengths in MPs was not found. Based on the QCL characteristics reviewed in section 2.1. and at the onset of this section we hypothesise that –at least, tentatively – the depth penetration values reported above for medical applications might be a proxy to figure out how far the QCL beam travels in the plastic particles. Clearly plastics are not soft tissue but, at the same time, they do not contain water, proteins or fat (strong mid-IR absorbers).

To gain some insight on this issue we present in the following paragraphs of this section a study dealing with LDIR spectra at different particle sizes and another using FTIR micro reflectance, to help interpret the results of the former. To the best of the authors' knowledge this is addressed first time in the MPs literature.

### 3.2.1. LDIR spectra and particle size

All the studies were made with more than 100 particles of each polymer and sizes varying from 20 to 1500 µm (section 2.3), deployed on a reflective Kevley slide without further treatment. The spectra of some particles were compared considering the intensities at the most intense peaks, as a function of the particle size. However, determining the particle size is anything but simple. The explanations below can be considered as a preliminary qualitative model because they assume that the height of the particle (the QCL-LDIR radiation illuminates the particles vertically) is directly related to the diameter of the particle and that this represents its size. In our studies we could observe that some particles were relatively spherical but that was not the case for all of them. Therefore, despite it is true that the system locates the highest/brightest point of each particle and directs the laser there, it has to be acknowledged that the correlation between height and diameter is not direct. So far, it is not possible in IR spectroscopy to measure the height of the particle whose spectrum is being acquired.

Results showed that there were four different behaviours, depending on the polymer. **Table 1** summarizes them and the regression equations (absorbance vs. size) for each polymer. For those materials with a double behaviour a regression equation is shown for each one.

To explain this empirical fact, recall **Figure 3a** and related explanations. Consider that the total absorbance a QCL beam experiences (at a given wavenumber) when passing through a particle is the sum of three terms: external reflectance at the incident point (transformed to absorbance as –$\log_{10}$(R)); absorbance when passing through the particle (the “transmission” term), and absorbance when reflected (the second term of transflectance). Internal dispersion and backscattering are also possible, but they will not be considered here although it is acknowledged that some plastics may contain clusters of compounds or crystalline phases and hence, they are not 100 % homogeneous. Note that all explanations below refer to a given wavenumber and the polymers employed in the study, as presented in section 2.3 (other formulations with more or different additives might lead to different effects, and this needs to be studied), which is not to be repeated in order to simplify writing. Four situations were observed:

1. For a totally reflecting material the first component should be the only relevant one (external reflection or reflectance) and would be more or less constant (but for diffuse reflection variations) and independent of the size of the particle. That seems the case for PVC (**Table 1**). It is true that the absorbance of the evanescent wave might be considered as well but that should not be relevant for the discussion here (as, indeed, for a given wavenumber the penetration due to this effect should be approximately constant regardless of the particle size).
2. A material that allows the passage (transmittance) of the radiation through its overall extent, without major absorption (think of this as an “ideal transflectance” situation) will show an increasing direct relation between total transflectance (absorption plus reflection) and penetration (strictly, height of the particle, which represents its size) because total transflectance increases with it (recall that all terms are transformed logarithmically to get absorbance-like values). That would be the case for HDPE (see **Table 1**, @1638 and 1304 $cm^{-1}$), PA, PET, PS, LDPE (@1304 $cm^{-1}$), and PP (@1102 $cm^{-1}$).
3. A polymer that absorbs the radiation (although not completely) will show only “good” transflectance for (very) small particles and it will follow the behaviour above up to a penetration depth where the reflected radiation cannot abandon the MPs, mostly because internal absorption. Hence the total absorbance decreases continuously with size. In the limit, the largest particles will exhibit essentially a reflectance behaviour, following Kirchoff’s law. That seemed to be the case for PMMA where, likely, the polymer presents a strong total absorbance, even for the smallest particles we had, around 20 µm and, so, only the decreasing pattern was observed. The “pure” reflectance region was obtained between 500 and 700 µm, where a plateau was reached.

4. The last situation is somehow like the previous one although with the material presenting moderate absorption and permiting transflectance only up to a maximum penetration or particle size. Here, the smallest particles will show an increasing trend for the total absorbance (they do have transflectance) but after a given size transflectance decreases, as in the previous case. Hence, a double behaviour will be seen when total absorbance increases up to a point when it starts decreasing. That would be the case for HDPE and LDPE, @1467 $cm^{-1}$, and PP (several bands).

We hypothesize that the point where the two behaviours coincide (in this case, around 250 µm) can be considered as the maximum depth where transflectance occurs (for that polymer and wavenumber). This value agrees very well with the path lengths reviewed above.

**Table 1**. Total absorbance behaviour as a function of the particle size (the ↑ icon denotes an increase in the total absorbance, otherwise, the ↓ icon will be displayed).

| **Polymer** | **Behaviour/s** (@ cm$^{-1}$, (particle size range): trend) | |
|---|---|---|
| PVC | No influence | 1435, 1256, 1327 cm$^{-1}$ (90-500 µm) |
| PA | Increase | 1145 cm$^{-1}$ (20-800 µm): y = 6e-05x + 0.0927<br>1739.5 cm$^{-1}$ (20-800 µm): y = 0.0005x + 0.0866<br>1300 cm$^{-1}$ (20-800 µm): y = 0.0002x + 0.0423 |
| PET | Increase | 1740 cm$^{-1}$ (20-800 µm): y = 0.0003x + 0.4379<br>1142 cm$^{-1}$ (20-800 µm): y = 0.0005x + 0.2117<br>1300 cm$^{-1}$ (20-800 µm): y = 0.0006x + 0.155 |
| PS | Increase | 1456 cm$^{-1}$ (20-800 µm): y = 0.0003x + 0.5057<br>1492 cm$^{-1}$ (20-800 µm): y = 0.0003x + 0.6109<br>1600 cm$^{-1}$ (20-800 µm): y = 0.0005x + 0.5226 |
| PMMA | Decrease | 1456 cm$^{-1}$ (50-700 µm): y = -0.0007x + 0.6166<br>1736 cm$^{-1}$ (50-700 µm): y = -0.0011x + 0.7489 |
| HDPE | Increase | 1304 cm$^{-1}$ (20-1000 µm): y=0.0002x+0.0614<br>1638 cm$^{-1}$ (20-1000 µm): y=0.0002x+0.0664 |
| | Increase+decrease | 1467 cm$^{-1}$ (<200 µm): ↑ y = 0.0033x - 0.0031<br>1467 cm$^{-1}$ (>200 µm): ↓ y = -0.0002x + 0.8401 |
| LDPE | Increase | 1304 cm$^{-1}$ (20-800 µm): y=0.0003x+0.046 |
| | Increase+decrease | 1467 cm$^{-1}$ (<200 µm): ↑ y = 0.0033x + 0.0325<br>1467 cm$^{-1}$ (>200 µm): ↓ y = -0.0004x + 0.9566 |
| PP | Increase | 1102 cm$^{-1}$ (20-800 µm): y = 0.0007x + 0.1014 |
| | Increase+decrease | 1460 cm$^{-1}$ (>200 µm): ↑y = 0.0041x - 0.0609<br>1467 cm$^{-1}$ (>200 µm): ↓ y = -0.0006x + 0.9336<br>1164.5 cm$^{-1}$ (<200 µm): ↑ y = 0.0028x - 0.0964<br>1164.5 cm$^{-1}$ (>200 µm): ↓ y = -0.0002x + 0.7971<br>1375.5 cm$^{-1}$ (<200 µm): ↑ y = 0.0031x - 0.0809<br>1375.5 cm$^{-1}$ (>200 µm): ↓y = -0.0007x + 0.6993 |
| PC | Increase+decrease | 1511 cm$^{-1}$ (<200 µm): ↑ y = 0.0043x - 0.0281<br>1511 cm$^{-1}$ (>200 µm): ↓ y = -0.001x + 1.0798<br>1267 cm$^{-1}$ (<200 µm): ↑ y = 0.0034x - 0.2173<br>1267 cm$^{-1}$ (>200 µm): ↓ y = -0.0004x + 0.6131<br>1600 cm$^{-1}$ (<200 µm): ↑ y = 0.0029x - 0.0652<br>1600 cm$^{-1}$ (>200 µm): ↓y = -0.0009x + 0.9141 |

To explain why some polymers (in particular, HDPE and LDPE) present different behaviours when different wavenumbers are considered we resource to the influence of the complex index of refraction on the appearance of infrared bands that involve transflection. This phenomenon is known since the early 1970s when Greenler et al. [61] studied band shifts and differences in band shapes between reflection-absorption spectra and traditional transmission ones. The variation of the index of refraction is more relevant for strong bands, which causes considerable changes from the usual transmission spectrum, but the transflectance spectrum of a weak band is much like the transmission spectrum. Further, it was suggested that changes on the bands in transflection measurements can be caused by reflections and interferences at the sample interfaces, including that between the sample and the reflective surface [48].

A consequence of these effects is that although the most intense spectral bands recorded by QCL-LDIR for a given polymer are present at any given particle size, the changes in their band intensities cause the general appearance of the spectra to vary with particle size (see **Figure 5** for an example). This justifies why it is so important that users update their databases considering spectra of different polymers at different sizes.

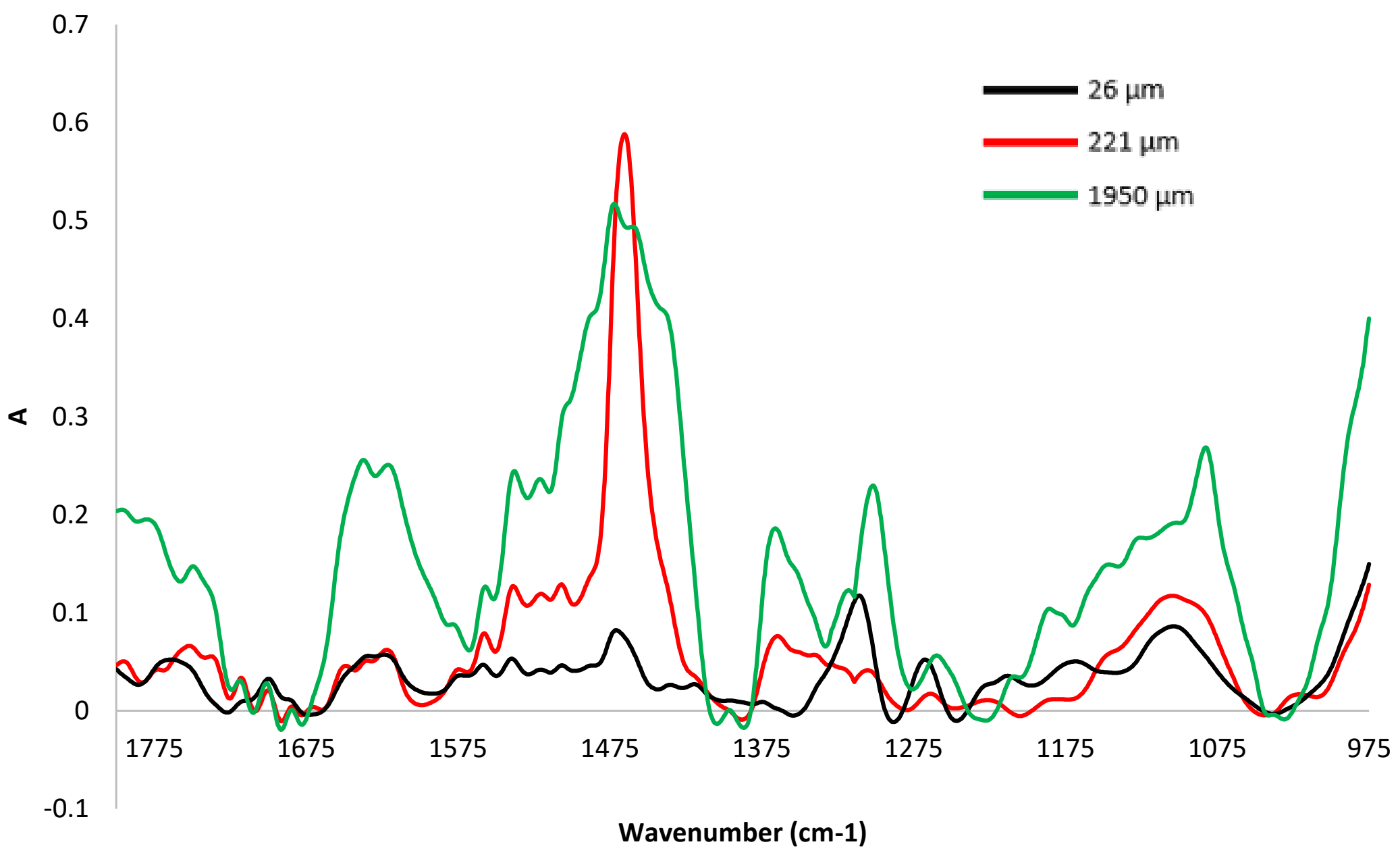

**Figure 5**: Spectra of pristine PE particles recorded by QCL-LDIR for different particle diameters: 26 µm, in black; 221 µm, in red; 1950 µm, in green. It is seen that the relative intensities of the bands change, but they are present in all cases.

### 3.2.2. Micro reflectance spectra and particle size

To compare the results obtained in the previous section with a standard technique, a similar study was undergone with micro reflectance FTIR (Globar source and MCT detector, see section 2.3). Only two polymers that showed a "double behaviour" in QCL-LDIR results were considered: PC and LDPE (@1467 $cm^{-1}$). In FTIR we could estimate the height of the particles manually using the focusing capabilities of the microscope. For each particle the supporting plate and the top of the particle were focused (the difference in the coordinates estimates the height). Unfortunately, those particles were not those where the LDIR spectra had been measured because they could not be inmovilized for transport and they could not be identified/marked. Both polymers revealed a linear increasing trend when the total absorbance was plotted against the "measured height" considering the same wavenumbers for which they had a double behaviour in QCL-LDIR (and roughly the same size range, see **Table 1**). It was also verified that the micro reflectance-FTIR spectra had less bands than their LDIR counterparts in the common ranges (see **Figure 6**). Therefore, it seems that the QCL produces some effects on the spectra that cause the micro reflectance (FTIR) and transflectance (LDIR) behaviour not to follow the same pattern. Note that for LDPE the micro reflectance peak around 1467 cm$^{-1}$ got distorted for largest particles, as indicated by the numbers in the legends of the subplots in **Figure 6** (we did not find a sound reason) and, also, the higher S/N ratio for the LDIR signals because in FTIR 50 scans were averaged.

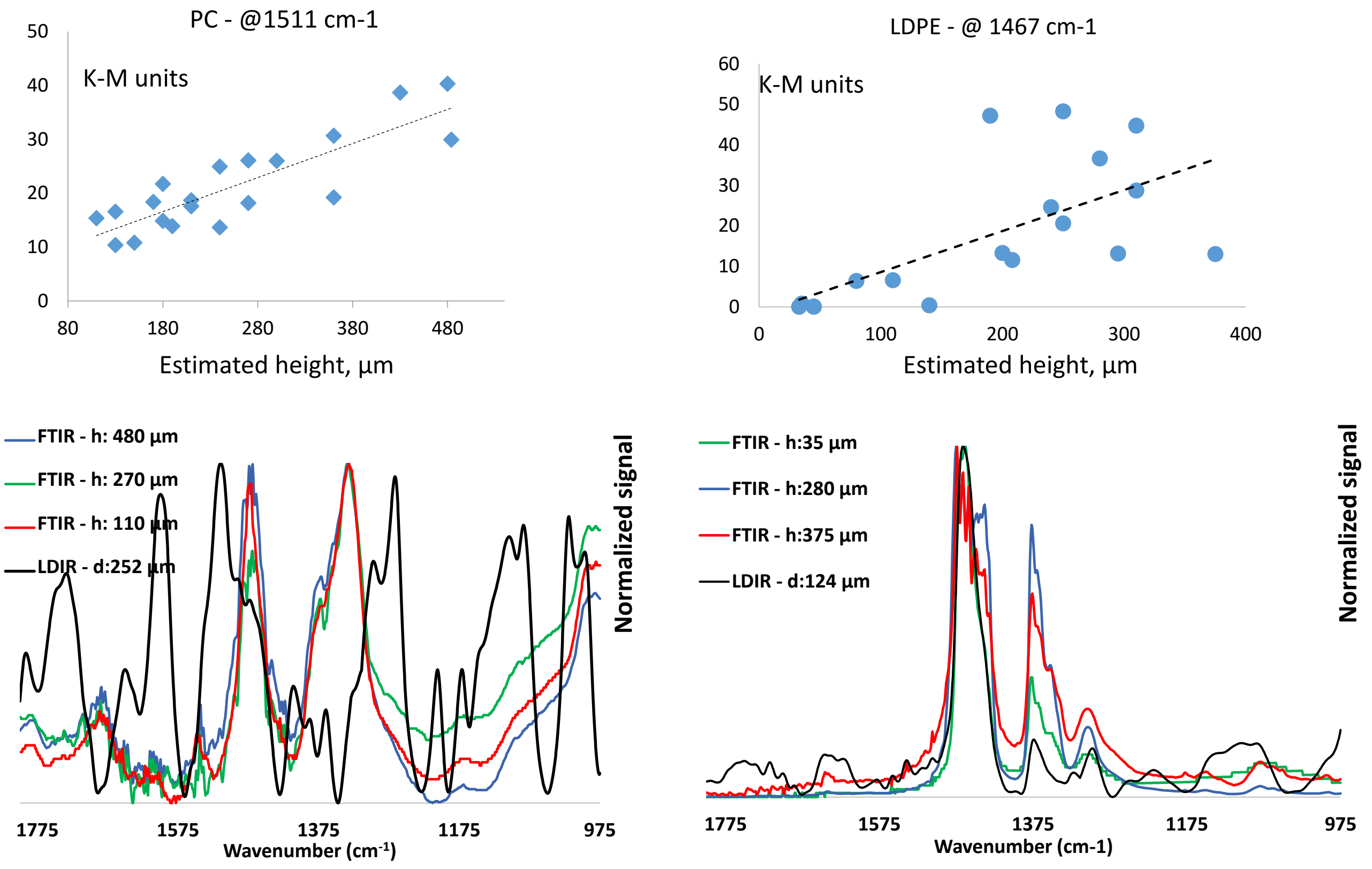

**Figure 6**: Evaluation of the behaviour of the total absorbance (Kubelka-Munk, K-M, units) when spectra of two polymers are recorded as a function of the estimated particle height (see text for details). The bottom figures show the micro reflectance spectra (FTIR) of three exemplary particles

(their estimated heights are shown in the legends, in µm), overlaid with characteristic LDIR ones. All spectra were normalized to unit height in the common spectral range.

The discussions above to justify the appearance of the LDIR spectra can be a reasonable starting point to explain the empirical facts, although they have several limitations. A potential one being the heating of the particles by the QCL beam at the incidence spot. This would modify the local density of the polymer and induce local changes on the bonds and structure. That would also contribute to changing/modifying the spectral bands. We acknowledge that this is a quite well-known problem in Raman. Nonetheless, in our view heating is not likely a major effect here because the lasers employed for Raman are much more energetic and impact longer over the particles (on the order of 30 s *vs.* >1 s in QCL-LDIR). Hence, total exposure to a wavenumber amounts milliseconds. To fully address in detail the extent of this question we would need to measure the temperature of the particles while registering the LDIR spectrum, but this is not possible. Further, the model might be refined if an instrumental device could measure the height of each particle (and temperature) during characterization, which is not at our disposal.

The very underlying cause of the behaviours observed in **Table 1** and **Figures 5 and 6** is the inherent chemical and physical nature of the polymeric particles as the molecular structure of the polymer, its density and inner linkages between the polymer chains determine how the radiation interacts with the sample. Indeed, the extent that light is transmitted, reflected, scattered or refracted depends upon the sample morphology, crystalline state, the angle of the incident light, and the difference in refractive indices of the sample and surrounding matrix, typically air [108]. It is also worth recalling (see section 2.2.) that the refractive index of a sample is a complex number with both a real and an imaginary component, the latter yielding a physical effect at some wavenumber known as "anomalous dispersion", which distorts the classical absorption band [48,61].

To complicate things, two further questions have to be mentioned:

(i) depending on the (polarised) wavenumber some peaks can appear more intense than others, or the other way around (compared to the "normal" transmission ones);

(ii) as noted in section 2.2., transflectance was developed initially to measure (very) thin films, in the order of nm or even Å [61] and in that case its physical grounds are well established [39,40]. On the contrary, much less papers were devoted to study the effects caused by "large" films or particles (recall the discussion in section 2.2. about the "big" size of MPs) although a practical paper reported different spectra as a function of the film size [65], as we do find here.

To conclude, QCL-IR measurements of small particles would typically exhibit a –essentially-transflectance mode while the bigger ones may be more influenced by other phenomena; mostly, absorbance. Hence, it can be said that more practical studies are needed to fully understand the behaviour of the particles under QCL-IR beams and try to interpret their spectra chemically. Right now, the most practical option to ameliorate this problem is to feed the spectral database used as reference with spectra of known polymer particles at different sizes.

These findings reinforce the need to expand the QCL-IR database with polymers of different sizes. This suggestion adds to ISO 24187:2023, where the importance of using specific libraries is mentioned [109], but it does not indicate that those databases should include different particle sizes. This might also be reflected in annex F (specific for QCL-IR) of the final version of the new ISO/DIS 16094-2 [97].

### 3.3. Instrument sensitivity

Section 2.4. showed that when the usual operational mode is employed ("particle analysis"), a topographical image based on the reflectance at a given wavenumber (e.g. 1442 $cm^{-1}$ to locate the C-H bending absorption) is obtained throughout the reflective plate. The particles are located based on contrasts of reflected light as they appear as bright illuminated pixels over a dark background.

The current controlling software allows selecting up to 11 levels of "sensitivity" (i.e., threshold levels to visualize bright pixels). Selecting a too high sensitivity (i.e., a too low contrast threshold, 9 or more) is a way to visualize smaller and fainter particles. Disappointingly, this may lead to the detection of some structures of the surface where the particles are deposited and the system analyses them ("false detections" of particles). This is depicted in **Figure 7**, where a small measuring region can be seen. Out of the around 18 potential "particles" (structures) only one corresponded to a real particle, as per visual and magnification observation. Such false detections can be noticed because their spectra show no characteristic features (they look like noise or the background) and will not match with known polymers. This is not highly relevant for MPs reporting but for indicating the total number of particles.

Those superficial structures can be due to several reasons: very tiny particles (dust, etc.) that are lower than 10 µm (and cannot be measured usually), reflections caused by some minor superficial irregularities of the reflecting surface itself, some effects of the residues of the solvents after evaporation, chemical degradation of the reflective material, the pore structure of the filters, etc. If their number is high the overall measurement time will increase greatly due to unproductive measuring and computing time, and the number of unidentified particles will augment. Further, be aware that when working with metal-coated filters and high sensitivities are selected there is a tendency that parts of the filters are confounded with “particles”. A way to avoid these false detections is to increase the threshold of the spectral match or hit quality index and/or reduce the instrumental sensitivity For instance, Potter recommended to use intermediate sensitivities as the default option [37].

A somehow opposed problem may occur whenever a low sensitivity is set. This clearly might render faulty detections because the contrast threshold was too high, and some particles are not “seen”. The problem is a bit more complex for fibers since they appear fragmented and might be counted as several fragments instead of just only one (see **Figure 7),** or might not be detected at all. This is not a specific issue of LDIR measurements, but of all systems featuring particle finding options and, so far, visual observation is the only way to know whether this occurs. However, that is a manual, tedious and time-consuming task. The same can be said with regards to clumped particles and/or fibers. Each aggregation will be counted as a unique entity.

Accordingly, an *ad-hoc* evaluation of the effects the level of sensitivity causes on the number of detected particles is advised. The user must optimize this measuring parameter for each kind of sample and treatment at hand, especially if the latter includes a final step where a solvent is evaporated over a Kevley slide. In our experience, intermediate values (5 or 6) yield usually good results.

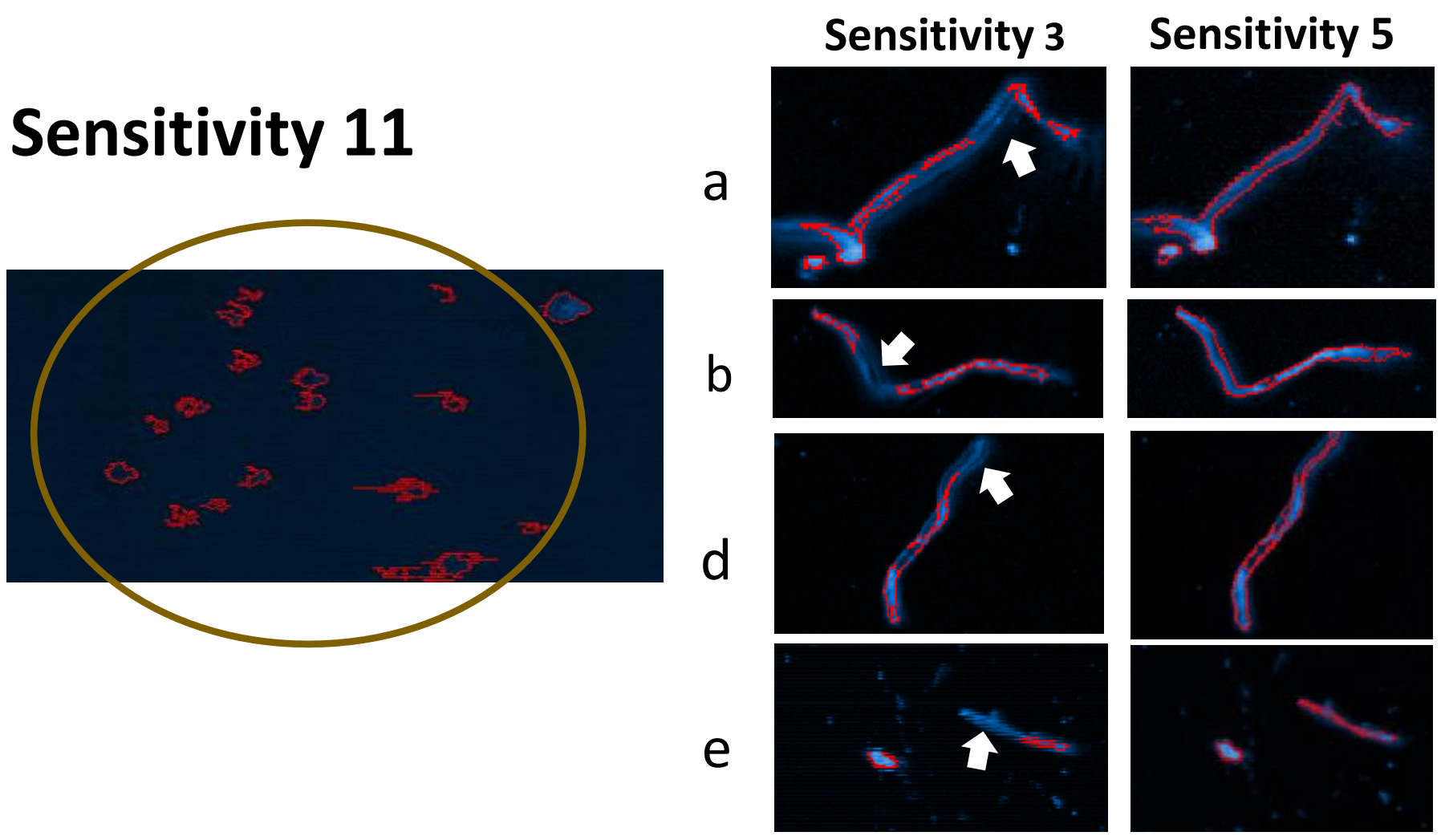

**Figure 7:** Effect of the sensitivity threshold level: false detections at too high sensitivities and incorrect detected shapes for fibers at too low sensitivities. White arrows: undetected shapes of fibres when sensivity = 3. For example, in (a) one fibre appears as 3 particles. The fibres shown were around 500 µm length x 10-20 µm width.

A related issue that merits some attention is the way in which the spatial resolution (defined as the minimum separation distance at which two particles can be differentiated) and the corresponding instrumental limit of detection (LOD, being this the capability of the instrument to detect isolated particles whose size is >LOD) can be enhanced in LDIR. This depends on many factors that are beyond the scope of this tutorial (for instance, the refraction index of the material, the diffraction limit of the radiation, the numerical apertures of the objective, the focusing capability of the light beam, etc.). Theoretically, the instrumental LOD of the LDIR can be set close to 6 µm when a one-particle-at-a-time working mode is selected (or for other applications, using the ATR device, thanks to the high refraction of its crystal). In current practice when automatic detection of the particles is set, this limit is higher, around 10-15 µm. However, how can it be improved to perform specific (manual) investigations?

First, it has to be indicated that there is only one optimal zoom setting for a particular combination of wavelength, numerical aperture, and objective magnification: the one that provides a pixel size in a digital detector that matches the Nyquist criterion (the scanning frequency conditions to register a signal correctly, either in space or time) [63]. Curiously, when operating at zoom settings of the microscope that are higher than that corresponding to the Nyquist condition, smaller "optical" pixels can be generated over the particle (i.e., smaller scanned areas) by oversampling the data within the reduced scanning area. This is useful to sample a specimen and obtain information about the position of a feature. It is worth noting that no details about the particle feature are provided beyond that allowed by the optimal conditions, but the location of the specimen can be determined best, provided that the particle is sampled by several detector pixels. The centroid of the spot identifies its location within the limits imposed by the S/N ratio [63]. This is a common approach applied even in FTIR [64].

Bear in mind that the diffraction limit of any optical microscope is a result of diffraction processes of light (as any physical objective has lense(s) and wall(s)) and the wave nature of light. The highest achievable point-to-point resolution that can be obtained with an optical microscope is governed by a fundamental set of physical laws that cannot be easily overcome by rational changes in objective lens or aperture design (a nice, general-purpose explanation can be seen elsewhere [110]). As a result of the diffraction limit, two MPs cannot be optically resolved if the distance between them is smaller than the diffraction limit for a specific setup (this is referred to as the Rayleigh criterion). Roughly, the diffraction limit is a bit less than half the wavenumber. For example, for the LDIR system, the diffraction limits at 1800 $cm^{-1}$ and 975 $cm^{-1}$ are 5.56 µm and 10.26 µm, respectively.

In a QCL-LDIR the IR beam is so focused and coherent that the spatial resolution is determined by the diffraction limit of the radiation not by the pixel size of the detector. The movement of the stage with the sample can be done in so small step sizes (even 1 µm), and the spectra are obtained so fast, that it is possible to oversample the particle, i.e., to measure the same particle with the focused beam spot slightly displaced a number of times depending on the pixel resolution to be obtained. Therefore, resolutions lower than the typical diffraction limit of 10 µm are possible. Spots can be overlaid as shown in **Figure 8**, thus improving the pixel resolution (and, so, the spatial resolution) by more accurately defining the shape of the particle. This option will not be exploited in most current applications but it can be of help for specific special-purpose searches (not automatized).

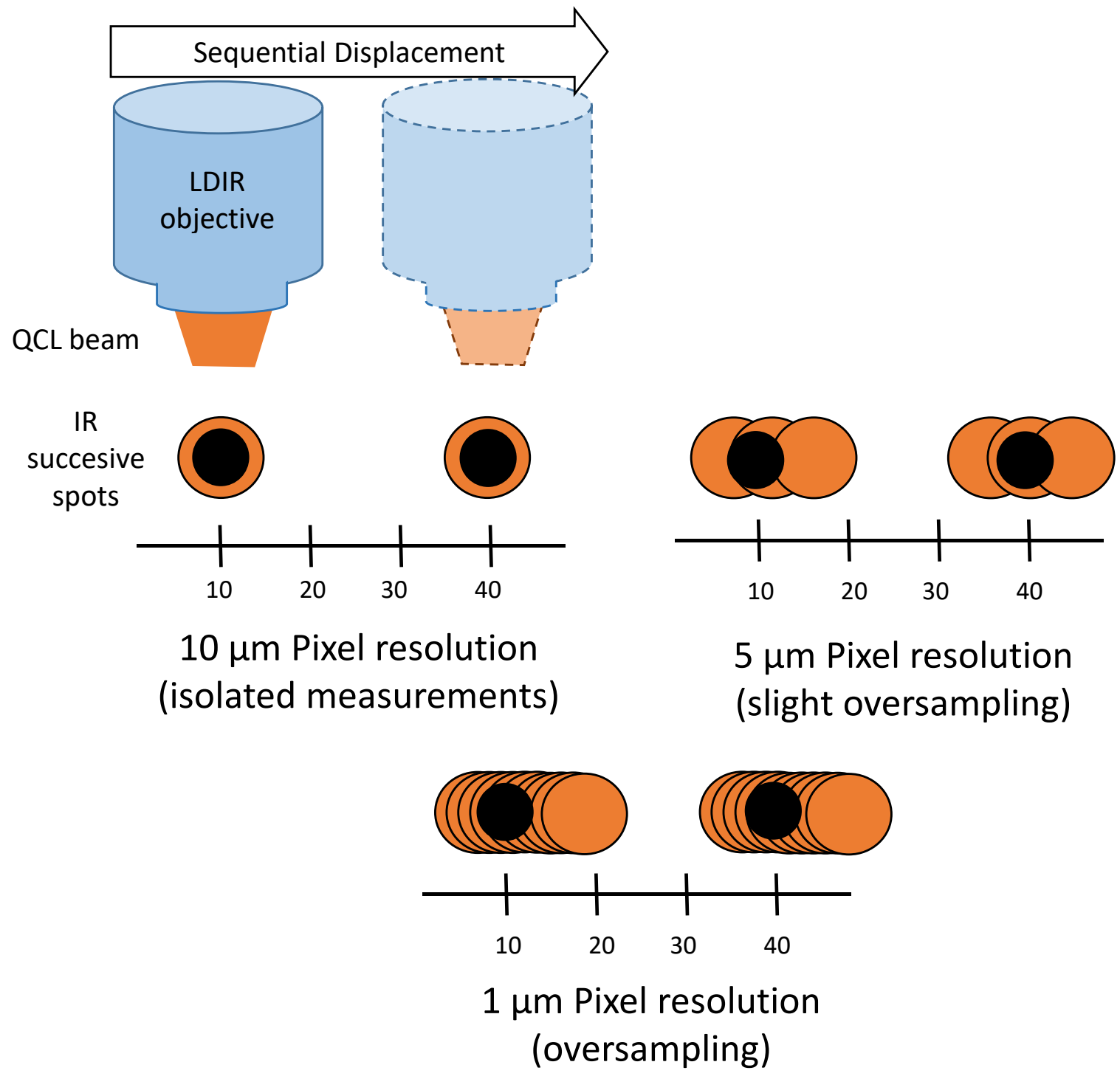

**Figure 8**: Improving the spatial resolution by oversampling a particle. The measurement spot has a constant size of 5.5 µm but better spatial resolutions are obtained by overimposing the spots. Oversampling allows for an increased spatial resolution and reduces the instrumental LOD as the particle is scanned like in “thinner” lines. In the example, the black particles (size <10 µm) might be undetected if isolated measurement IR spots are used, oversampling allows to locate them.

### 3.1. Measurement and processing speed

The overall speed of the LDIR system depends on the number of analysed particles and photographs and spectra accumulated at the computer memory (as outlined in section 2.4.). To understand this recall that a first detection of all particles in a selected area is done; then for each particle, a photograph is taken (if requested) and its spectrum recorded. A comparison with the database is done and a series of match indexes calculated. All this information is stored in a same electronic file before moving to the next particle. The more particles are analysed, the more time the computer requires to manage each entry. This issue is relevant to plan the laboratory workflow.

The LDIR can measure 600-700 or 200-300 particles per hour, depending on the definition of the sampling areas and number of particles accumulated in the computer memory. To test this effect three extensive trials were made. A sample with a large number of particles was selected and several strategies to define the areas of study were set. Then, a total measuring time of 24 h was established. The photographs were disabled to save time and the minimum size of the particles to be characterized was limited to 10 µm. The instrumental settings were the same for the three experiments but for the size of the regions to be scanned.

When the entire plate was considered as a unique sampling area (**Figure 9a**, 1656 $mm^2$) 15,889 particles were detected and only 3,338 ones were measured in the first 24 h. When the sampling area was halved (**Figure 9b**, 840 $mm^2$) 16,081 particles were detected and 8141 ones analysed in the first 24 h. When the sample was divided in small areas (**Figure 9c**, around 41.2 $mm^2$/each) 18,012 particles were detected, and 9,287 ones studied in the first 24 h. Velocity was not constant as it decreased steadily. At first, 582 particles/h (4-5 s/particle is the standard speed) were measured and after 24 h the speed was 303 particles/h (18 s/particle). These results agree well with Potter [37] who set that the lowest analysis time was achieved with the default sensitivity and fast scan and sweep speed.

Note that the number of particles detected is not the same when different strategies to “divide” the reflecting plate before the measurements are established. Strategy c (**Figure 9c**) allows the detection of more particles in total. This appeared more remarkable for gold-coated filters than for Kevley slides (**Figures SM3 and SM4**, Supplementary Material). This is because the IR imaging gets more contrast when small areas are selected. So we recommend defining small areas (e.g., around 8x8 mm) but avoiding placing fibres between two subregions.

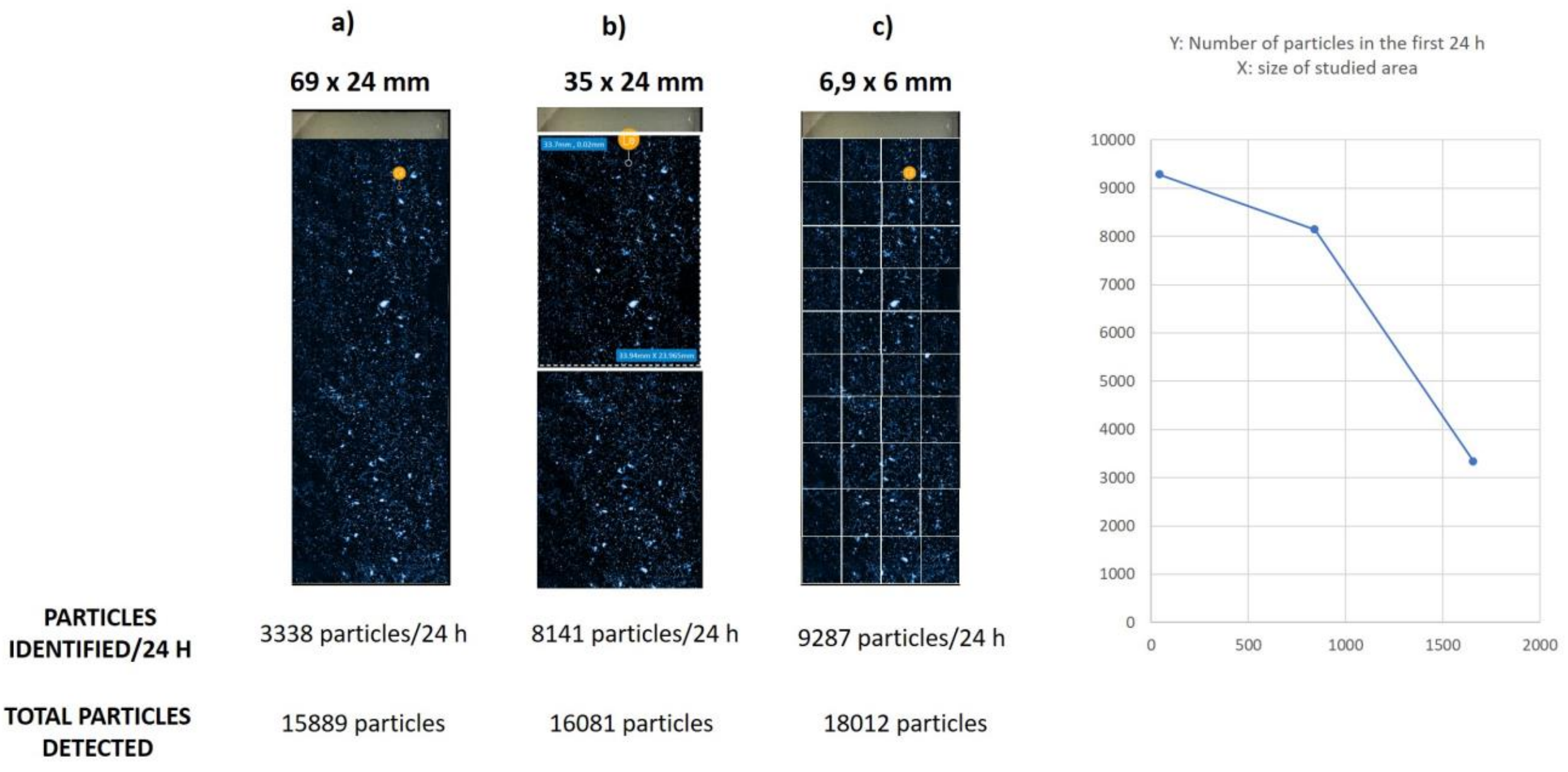

**Figure 9:** Effect of the size of the measuring area on the speed of the overall measurement. The rightmost subplot depicts the reduction in the overall measuring speed when the measuring area increases.

Setting adequate lower limits to address the particles to be measured in each particular problem is not trivial and will affect the laboratory sample throughput. This is important for industrial or environmental monitoring purposes. In particular if each national government regulates the size of the particles for monitoring studies. If the limit is set too low; e.g., below or around 10 µm, the number of particles may increase exponentially the measurement time. It is quite useless to measure only a sample per week as it indeed happens nowadays in some research studies that must study complex field samples (and recall that this adds to the usually long sample treatment time).

To overcome this problem some authors suggested to measure only a part of the filters throughout which the samples were filtered. Although this has been proposed mostly for Raman and FTIR [111–113]. Two recent papers presented preliminary approaches for subsampling in QCL-LDIR [27,114]. However, one has to be aware that any subsampling will yield relevant errors, around 25-30 %, that although important might be acceptable due to the inherent difficulty of the samples (mostly whenever as large number of particles as 30,000 have to be studied in a Kevley slide; e.g. from soils). This issue deserves further study and development but it might be a reasonable alternative to increase laboratory throughputs and reduce costs.

### 3.2. Hit quality index (HQI)

A critical step of the chemical characterization process is to identify the polymeric nature of the particles, which is done by matching the obtained spectrum with the library/ies. A parameter to quantify the correlation or match; i.e., how close a spectrum is to another is the “Hit Quality Index”, HQI. This index is analogous to the “correlation” or “correlation index” used in other programs. In this case it is the cosine of the similarity, rescaled to a 0-1 scale. Similarity was defined as the dot product of the first derivatives of the measured spectrum and that of the database, normalized by their modules. In addition to the fact that there can be many mathematical options to carry out such a comparison, this is a challenging step because a unique, objective threshold cannot be established due to the use of different libraries, the different experimental conditions, physical and chemical state of the polymer, etc. While the LDIR integrates a library of polymers and other common particles the user is highly advised to create or enlarge existing libraries with their own spectra and polymers typical of their particular field of work, also with different presentations (fine powder, particles, little pellets, weathering levels, etc.) including the result of the sample treatments undergone to get the final measuring specimens. This is an excellent way to improve the matching capabilities since a custom-built database will represent best his/her real measurement conditions, as well as resemble best the general aspect of their spectra. The practical application of this strategy will not take much time because comparing the already measured spectra against different databases with the software is not really a problem. Considering various libraries for particle identification is straightforward as the software allows for simple reprocessing of the already recorded spectra (no further measurements again), this is illustrated in **Figure SM2(b)**, Supplementary Material.

So far, a wide variety of threshold levels to accept a match as a “good” identification was considered in LDIR-related literature (**Table 2**). In our view, many of them are too low because the QCL-based transflectance spectra are very rich in number of peaks and, so, medium or even high HQIs can be obtained easily, despite the real overlap between the spectra is poor. In total, 74 articles were found were the HQI was mentioned. In 18 papers the threshold HQI is not disclosed, which is not good practice. Most of the remaining works established HQI thresholds < 0.85 (even of 0.6). In our experience, this might involve too many false positives and poor identifications when field samples are considered, unless they are validated as per visual scrutiny of the analyst. However, it also depends on the spectra included in the library, as it will be discussed in subsequent sections.

**Figure 10** exemplifies a comparison of a spectrum of a known type of fiber with the default library and a custom-built one. Somebody might be happy when they find an HQI of around 0.7 and will inform that a polyurethane particle had been found (**Figure 10**, up). However, a mere visual observation of the spectra reveals that this is not so and such a “match” should not be reported. A custom-built database revealed a nice and sound match with weathered PET (at it was indeed the case), with a much better HQI (0.95, **Figure 10**, bottom) and reasonable spectral overlap. The matching step may be cumbersome as some matches with even high correlation indexes (or HQI) may not correspond truly to the proposed polymer, as shown in this example. Therefore, we would recommend to be conservative in order to secure subsequent decision-making (policy-makers, regulations, etc.). This is a decision that any researcher has to make but it should be stated clearly in the reports.

**Table 2.** Literature review of the threshold HQI criteria found in literature that employed the QCL-LDIR measurement system.

| HQI suggested | Matrix | Study |
|---|---|---|
| **Not disclosed** | Sediment | Bringer et. al., (2021)[115] |
| | Soil | Li et al., (2021)[29] |
| | Stormwater | Olesen et al., (2019) [116] |
| | Groundwater | Samandra et al., (2022)[117] |
| | River water | Bauerlein et. al., (2023)[83] |
| | Freshwater | Scircle et al., (2020)[76] |
| | Textile wastewater | Li et al., (2023)[118] |
| | Indoor house dust | Lim et al., (2022)[119] |
| | Human faeces | Zhang et al., (2022)[120] |
| | Estuary water | Cheng et al., (2021)[121] |
| | River water | Cizdziel, (2020)[122] |
| | Soil and sediment | Hao et al., (2021)[123] |
| | Surface freshwater | Mughini-Gras et al., (2021)[84] |
| | Bottled water | Song et al., (2021)[124] |
| | Atmospheric | Liu et al., (2024)[125] |
| | Endometrial polyps | He et al., (2025)[126] |
| | Soil | Kang et al., (2024)[127] |
| | Coastal air | Liu et al., (2025)[128] |
| **0.60** | Wastewater | Kools et al., (2021)[129] |
| | Anti-corrosion coatings | Hildebrandt et al., (2024)[130] |
| **0.65** | Sediment | Cheng et al., (2021)[27] |
| | Wastewater | Tian et al., (2021)[77] |
| | Septic tank | N. Liu et al., (2022)[85] |
| | Human Sputum | Huang et al., (2022)[88] |
| | Sediment and surface water sample | Wu et al., (2023)[131] |
| | Indoor dust | Peng et al., (2022)[91] |
| | Petrochemical Sludge | Deng et al., (2023)[132] |
| | Agricultural Soils | Jia et al., (2022)[89] |
| | Placenta | Liu et al., (2022)[133] |
| | Mineralized refuse leachate | Zhang et al., (2021)[134] |
| | Bird lungs | Wang et al., (2025)[135] |
| | Gallstones | Zhang et al., (2024)[136] |
| | Soil | Lotz et al., (2024)[137] |
| | Soil | Wang et al., (2024)[138] |
| | Atmospheric deposition | Zheng et al., (2024)[139] |
| | Sediment | He et al., (2024)[140] |
| | Water | Yang et al., (2024)[141] |
| **0.7** | Surface water | Whiting et al., (2022)[35] |
| | Oysters and Mussels | Ribeiro et al., (2023) [92] |
| | Cephalopods | Ferreira et al., (2022) [142] |
| | Urban river water | Fan et al., (2022) [143] |
| | Bakeware (using QCL-based microscope) | Lin et al., (2024) [144] |
| | Pig lungs | Li et al., (2023) [30] |
| | Landfill leachate and groundwater | Wan et al., (2022) [145] |
| | Oysters | Li et al., 2024 [146] |
| | Sun sponge | Ribeiro et al., (2024)[147] |
| | River sample | Chen et al., (2024)[148] |
| | Sediment | Yu et al., (2024)[149] |
| | Soil/fertilizer | Liu et al., (2024)[150] |
| **0.75** | Sediment, seawater, mussels and fish | Ourgaud et al., (2022)[26] |
| | Trail surface | Forster et al., (2023)[151] |
| | Atmospheric (rural & urban) | Lavoie-Bernstein, (2024)[152] |
| | Soil | Li et al., (2024)[153] |
| **0.8** | Bronchoalveolar fluid | Qiu et al., (2023)[154] |
| | Atmospheric dust | Liu et al., (2022)[90] |
| | Subsurface water | Hildebrandt et al., (2022)[155] |
| | Artic fjord water | Bao et al., (2022) [156] |
| | Sediment and marine sample | Ghanadi et al., (2023)[157] |
| | Oysters | Lin et al., (2024)[158] |
| | Maternal blood, fetal appendages, umbilical vein blood | Sun et al., (2024)[159] |
| | Soil | Shi et al., (2024)[160] |
| | Blood | Liu et al., (2024)[161] |
| | Longnose lancetfish | Gao et al., (2024)[162] |
| **0.85** | Surface water, sediment, soil | Qian et al., (2023)[163] |
| | Sold bottled water | Li et al., (2023)[164] |
| | Snails | An et al., (2022)[165] |
| | Seafood | Süssmann et al., (2024)[166] |
| | Salt | Gao et al., (2024)[167] |
| **0.9** | Semen | Zhao et al., (2023) |
| | Human Placenta | Zhu et al., (2023)[169] |
| | Surface water | Hansen et al., (2023)[170] |
| | Bulk atmospheric deposition | López-Rosales, (2023)[171] |
| | Mussels | Garcíaet al., (2024)[95] |
| | Foliage samples | Falakdin et al., (2024)[172] |

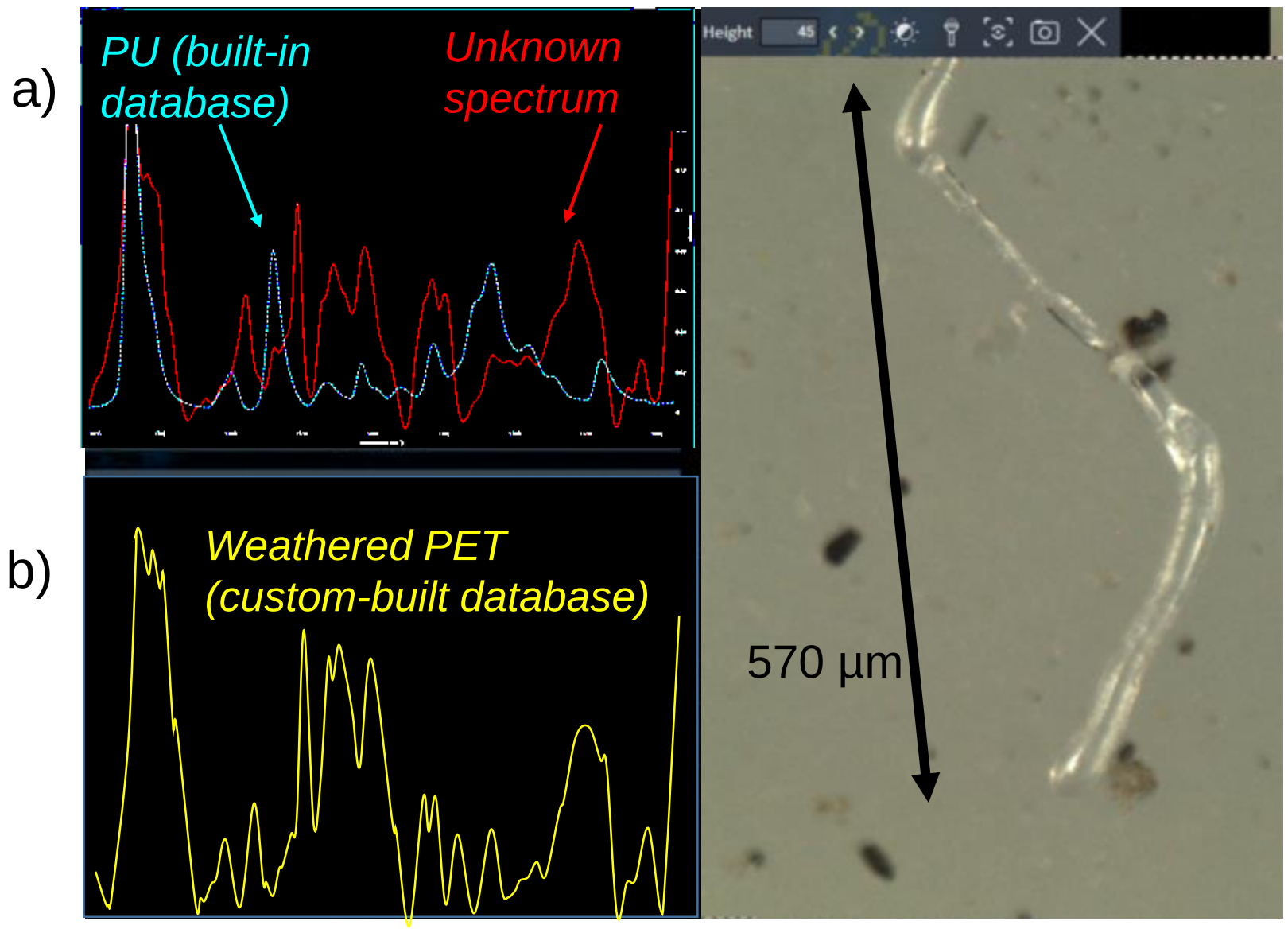

**Figure 10**: Comparison of the IR spectrum of a fiber) against the built-in library (upper subplot) and a custom-built database (lower subplot, see text for details).

Therefore, it is important to examine critically –as per visual observation (following traditional well-known good laboratory practices in IR spectroscopy)- the highest matches to be sure that the assignations are reasonable. A match index *alone* is not enough to ensure sound decision-making. Another interesting approach would be to apply other searching algorithms to assess the final findings, although this is not always possible because of the built-in options of the controlling software of the commercial instruments.

We acknowledge that the match indexes shown in this paper are higher than those currently observed in many published papers and this stems from attempts to solve many false assignations we found in our work (among them, that polymer weathering is not considered in current databases). The HQI values we use agree with those suggested by Potter [37]. Therefore, we preferred to restrict the number of false positives in our identifications.

A pragmatic suggestion that might address this issue is to set a criterion that establishes different levels of confidence for the spectral match (in any case, keep in mind that the scientist is responsible for verifying the adequacy of the stated matches, according to international quality guidelines like the European Directive 2004/10, about good laboratory practices). Thus, low, medium and high confidence HQI ranges can be set. **Figure 11** illustrates several real examples for common polymers. The reason why values <0.8 are not considered here is because after participation in two interlaboratory exercises those levels were too unreliable. In some cases, we experienced wrong assignations (noticed as per visual observation of the spectra) even when 0.85 < HQI < 0.90. Finally, after many visual comparisons of spectra we decided to set medium “confidence levels” to matches with HQI threshold $\geq$ 0.9 [171], like Potter [37] did. Recently, a comparison of results obtained by LDIR and Py-GC-MS stablished that an HQI of 0.85 yielded the most similar values [167].

What is also a good solution in the case of wrong assignments is to evaluate more manually the particle by a detailed line scanning or individual focusing (high magnification) and accumulating more spectra. If possible it is also good practice to compare the LDIR spectrum against other databases after a specific analysis (FTIR/Raman) to confirm its nature (although this has the problem of the shape of the spectra, as discussed above). Once its identity is confirmed add it to your LDIR library

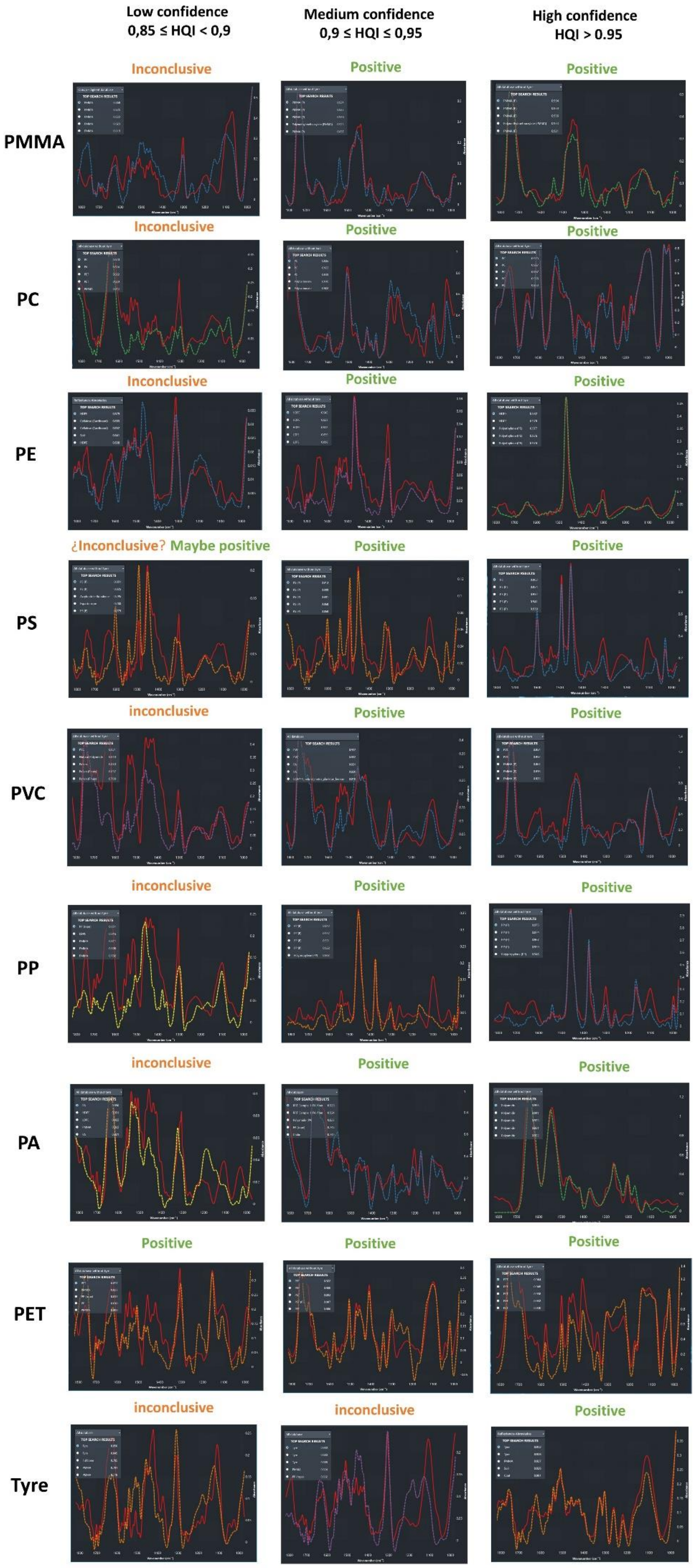

**Figure 11:** Match or hit quality index structured in three "levels of confidence": low, medium and high. The dashed lines correspond to spectra in the library.

Nevertheless, an exception was required for particles of tyres and rubber. The large number of peaks in their spectra leads to quite high correlations even when the assignations have no sense (as per visual validation). This is problem occurs when using FTIR and Raman as well. Following, we only accept assignations as tyres and rubber when HQI> 0.95, see **Figure 11** for an example [173]. In this sense a specific library for tyres was constructed including tyres from different brands and/or weathering stages. This agrees with recent ISO 24187:2023, which recommends HQI thresholds of about 0.8 for FTIR. However this cannot be extrapolated immediately to LDIR because of the specificities of LDIR spectra [109], this issue is commented also in ISO/DIS 16094-2 [97].

### 3.3. Fiber or fragment

Many studies need to differentiate between fragments and fibers. So far, this classification is not provided automatically but the software calculates several parameters useful for this purpose. In effect, the standard Excel-type file that can be exported after the measurement process contains four "geometric" indices, scaled from 0 to 1: eccentricity, solidity, circularity and aspect ratio. We will consider the last three. Solidity corresponds to the "ratio of the particle area over the area of its convex hull". A particle that occupies the shape of a rectangle will have a solidity close to 1 whereas a fiber will have a low solidity since its area is small relative to its bounding contour, more details can be found elsewhere [174]. A perfectly circular particle will have a circularity of 1; a fiber, close to 0. For example, Liu et al [85] used a circularity >0.6 to classify granules and fibers. The aspect ratio relates length to width and if >3 the fragment could be considered as a fiber, according to ECHA [6]. Nevertheless, recall that the sense of length and width can be interchanged depending on the orientation found for the fragment [35] and, so, an aspect ratio < 1/3 (0.33) would also be a fiber.

The aspect ratio must not be the unique parameter considered to detect fibers. In particular because thin fibers can have low aspect ratios if they are curved, as it is usual (see thin fibers in **Figure 12**). In this case an important complementary criterion to detect curved and tiny fibers is the solidity parameter. Particles with solidity <0.35 are fibers, while those with solidity >0.35 may be either fragments or fibers, because there are fibers that cover a high bounding area (coarse fibers). Following, we combined circularity and solidity**:** particles with solidity> 0.35 and circularity > 0.3 may be a “cylinder” or a round fragment, but if circularity <0.3 they may be coarse fibers or irregular fragments, depending on their aspect ratio.

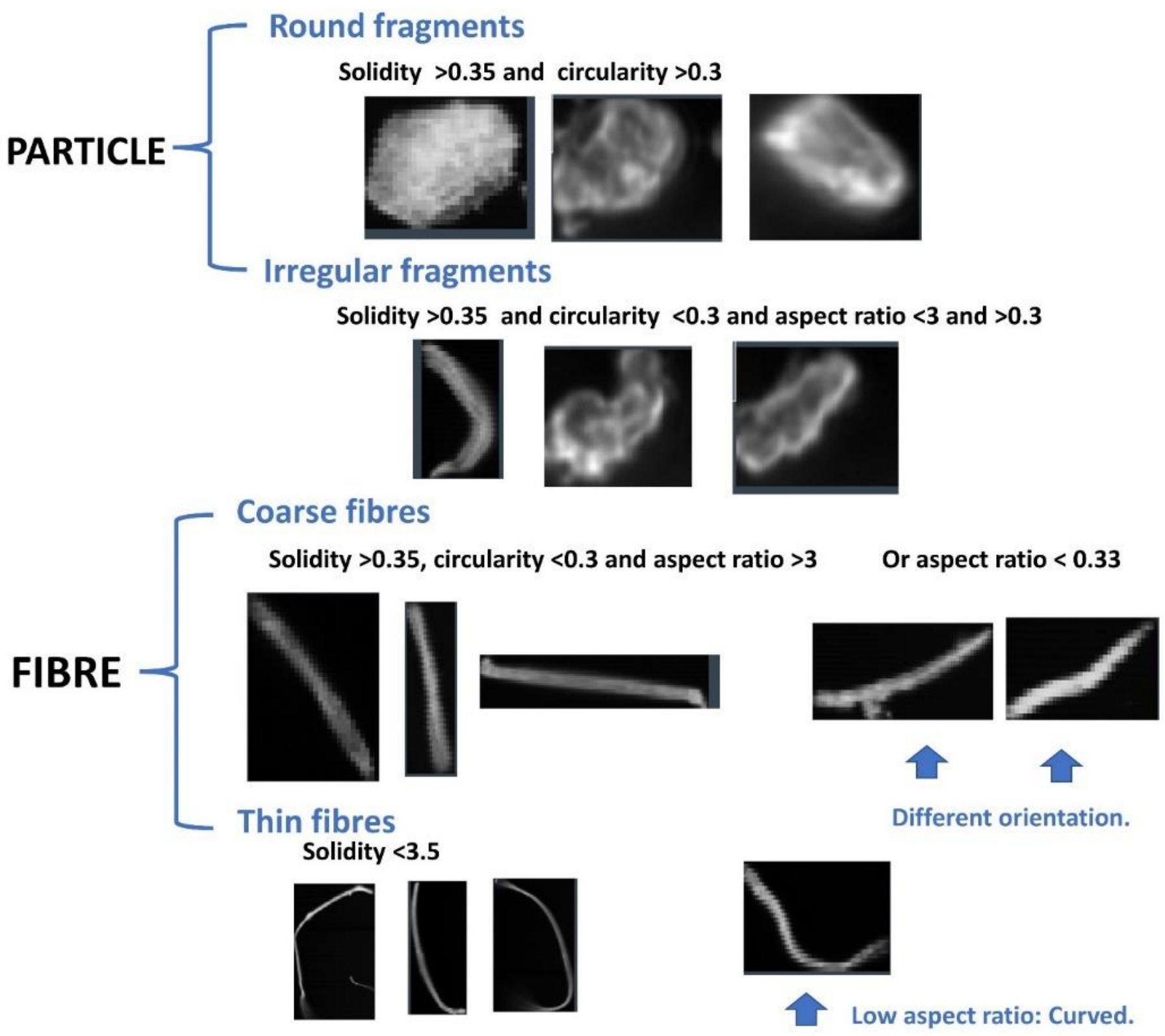

**Figure 12:** Tiered criterion to differentiate fragments and fibers according to solidity, circularity and aspect ratio.

**Table 3**. Tiered approach to differentiate fibers and fragments. (*) means that a coarse fiber can have an aspect ratio <0.33 or >3, depending on its orientation.

| Geometric indices calculated by the controlling software | | | Shape | Fragment or fiber |
|---|---|---|---|---|
| Solidity ≥ 0.35 | Circularity ≥ 0.3 | | Rolled fragment | Fragment |
| | Circularity < 0.3 | Aspect ratio ≥ 3 | Coarse fiber | Fiber |
| | | Aspect ratio < 3 and ≥ 0.33 | Irregular fragment | Fragment |
| | | Aspect ratio ≤ 0.3 | Coarse fiber* | Fiber |
| Solidity < 0.35 | | | Tiny Fiber | Fiber |

Thus, the tiered classification rule shown in **Table 3** was validated using 500 particles whose assignations were checked visually, where from a contingency table yielded 94 % success in fibers, and 93 % in fragments. This criterion was used in our previous works, see for example (López-Rosales et al., 2024) [171] and Falakdin [172]. This tiered classification criterion was implemented in an App developed by the Marine Environmental Studies Laboratory of IAEA (International Atomic Energy Acency) in Monaco, under the NUTEC Plastic research program. This App (called YABE) will include processing tools for exported LDIR data files and will be freely available soon.

## 4. CONCLUSIONS AND FUTURE TRENDS

In this paper an overview of the most relevant grounds and working parameters of a state-of-the-art QCL-based infrared analytical system was presented, called LDIR. Some phenomena can alter the effectiveness of the instrument and influence on the quality of the recorded spectra and, thus, on the final identification of putative MPs. To the best of the author's knowledge this is the only publication dealing with a technology used to characterize MPs that treats in detail all relevant aspects that determine its final outcome. Several questions presented in the chapters remain open and might be the subject of future studies, like studying how the spectra of different polymers become affected by transflectance phenomena, the study of the depth penetration of the laser beam into the particles, to set general-use sensitivity and HQI thresholds, how to set subsampling approaches, etc.

An interesting proposal for the LDIR community would be to perform interlaboratory comparisons where LDIR users compare their systems using one or several common samples. This would be a nice way to infer how different instrumental setups impact on the final outcomes (number and identification of MPs).

The system is able to measure around 500-600 particles/h (which opens an interesting way for monitoring) although that depends on the size of the scanning area and on the total amount (and lowest size) of particles to be studied. This is especially cumbersome whenever big amounts of particles are present on the reflective surface (e.g., >20,000). Our recommendation is to structure the work in several projects (containing no more than 7,000-8,000 particles each) and select various small measurement areas per project so that the computer optimizes the processing capabilities. An issue that deserves more attention in a near future is to set subsampling areas within the filters that are studied by QCL-LDIR. Despite this might yield some errors, literature reported already approaches that may lead to mature solutions for routine and high laboratory sample throughputs and reduce costs. This is of most importance for monitoring laboratories (mostly governmental, but not only) as nowadays the workloads are impractical for routine laboratories.

The theoretical grounds of reflectance-absorbance measurements for typical plastic particles (unlike thin films) is somehow underdeveloped from a practical viewpoint. Their spectra may be influenced by other phenomena than current transflectance; like the absorbance of the material. It was shown that, according to traditional criteria, the particles measured by LDIR can be considered as "big" when compared to the IR radiation. Besides, it was seen that the intensity of the spectra depends on the size of the particle, although not linearly (due to the chemical nature of the polymer and its complex index of refraction). Four behaviours were seen:

i. For a totally reflecting material, surface reflection would be the only relevant phenomenon and the total absorbance would be approximately constant regardless of the size of the particle.
ii. A material with low absorbance that allows transmission of the laser radiation will show an incremental positive relation between total absorbance and penetration depth (in the studied size ranges). This would resemble the "typical" transflectance behaviour.
iii. Polymers with high absorbance (although not 100 % absorbers, as in case i) will show transflectance for small or very small particles, and total absorbance will increase with particle size up to a point where the total absorbance decreases (no transflectance occurs) or reach a plateau.
iv. If a polymer absorbs the radiation moderately transflectance and total absorbance increase initially with particle size (like in ii), although after a maximum penetration depth (size) transflectance becomes weak and the total absorbance may decrease.

The intersection of the two behaviours observed in options iii and iv is hypothesized to be the maximum depth where transflectance occurs (for that polymer and wavenumber). That value was around 250 µm for some polymers, which agrees very well with other path lengths reviewed from literature, although in other research fields.

False detections of particles can occur due to several reasons. They reduce the overall speed of the analysis because of unproductive measuring and computing times. A way to avoid them is to increase the threshold of the hit quality index and/or reduce the instrumental sensitivity (i.e. the light threshold contrast of adjacent pixels in the image); in general, intermediate levels are suggested for the latter.

Finally, a literature review demonstrated that despite the reduced working spectral range of the QCL-LDIR it provides information enough to identify and quantify different analytes in several types of applications, and to differentiate among the most common polymers, albeit its information may have to be combined to chemometric tools (pattern recognition, clustering and classification, machine learning, etc.) to get best results. Further, three issues are of most relevance regarding chemical identification of putative MPs:

(i) Many of the so far proposed match (or hit quality) indexes are too low because the QCL-based reflectance spectra are very rich in number of peaks and, so, medium or even high HQIs can be obtained easily, despite the real overlap between the spectra is poor. The mere acceptance of the match proposed by the software is a bad practice; in particular, if the index is relatively low (e.g., <0.85). We propose to set a criterion that establishes three levels of confidence: low (0.85<HQI<0.9), medium (0.9<HQI<0.95) and high (HQI>0.95), and this agrees with current trends. A relevant exception is that we only accept matches for tyres and rubber when HQI>0.95. It is also a good practice to check visually the highest matches to ensure sound decision-making.

(ii) A tiered classification rule to differentiate fragments and fibers was set considering three geometric parameters: solidity, circularity and aspect ratio.

(iii) Finally, it is of paramount importance to bear it in mind that the user of a QCL-LDIR instrument should update the spectral libraries with their own spectra as that will resemble better his/her processing stages, the kind of polymers they measure, different size, degree of weathering, etc. so there is less chance for mismatches occurring due to different effects.

**ACKNOWLEDGEMENTS**

This work is a part of the European Union H2020-funded LABPLAS project (Grant No. 101003954). The Spanish Research Agency and the Ministry of Science and Innovation are also acknowledged (SplashMare project, Grant PID2022-138421OB-C21). The Program "Consolidación e Estructuración de Unidades de Investigación Competitivas" of the Galician Government (Xunta de Galicia) is also acknowledged (Grant ED431C 2021/56).

**DISCLAIMER**

All opinions written in this paper correspond solely to the authors' views and experience and none of them can be endorsed, not even partially, to Agilent Technologies.

# REVIEWING THE FUNDAMENTALS AND BEST PRACTICES TO CHARACTERIZE MICROPLASTICS USING STATE-OF-THE-ART QUANTUM-CASCADE LASER REFLECTANCE-ABSORBANCE SPECTROSCOPY

Adrián López-Rosales[1], Borja Ferreiro[1], Jose M. Andrade[1]*, Andreas Kerstan[2], Darren Robey[3], Soledad Muniategui[1]

[1]Grupo Química Analítica Aplicada (QANAP), Instituto Universitario de Medio Ambiente (IUMA), Universidade da Coruña, Campus da Zapateira, E-15071, A Coruña, Spain

[2]Agilent Technologies, Hewlett-Packard Str. 8, 76337, Waldbronn, Germany

[3]Agilent Technologies, 679 Springvale Rd, Mulgrave, VIC, 3170, Australia

# SUPPLEMENTARY MATERIAL

## 1.- GLOSSARY OF SOME TECHNICAL TERMS USED IN THE PAPER

**Absorbance**: A parameter defined by the Lambert-Beer-Bouguer Law, as $A= -\log_{10}(I/I_0)$ (decadic absorbance), where A is the absorbance, and I denotes the intensity of the radiation either after placing a sample specimen in the path of the sample beam or before that ($I_0$).

**Absorption (or extinction) coefficient (k):** Linear decadic (a,K) and Napierian absorption coefficients (α) are equal to the corresponding absorbances divided by the optical path length through the sample. The molar absorption coefficients (decadic ε, Napierian κ) are the linear absorption coefficients divided by the amount concentration.

**Attenuated total reflectance (ATR):** Spectroscopic technique based on the behaviour of the radiation when it hits an interphase between two materials (one of them would be the sample and another a highly reflective crystal in the instrument). If the incidence angle is adequate, the radiation experiences a total internal reflection (radiation stays inside the crystal) on the crystal´s surface, so only a small fraction of the radiation (*evanescent wave*) interacts with the sample.

**Brightness** (“brilliance” or radiant intensity), defined as the photon flux or power emitted per source area and solid angle -watt per steradian units- (it is confused often with power)

**Collimation:** The limiting of a beam of radiation to the required dimensions and angular spread.

**Conduction band:** A vacant or only partially occupied set of many closely spaced electronic levels resulting from a large number of atoms forming a system in which the electrons can move freely or nearly so. This term is usually used to describe the properties of metals and semiconductors.

**De Broglie´s wavelengths:** De Broglie proposed that, as photons have particle-like characteristics (mass and momentum and energy), particles in motion should have wave-like properties. Thus, the de Broglie wavelength is the wavelength associated to a specific particle, and is dependent on its mass and velocity.

**Decadic absorbance, $A_{10}$:** The negative decadic logarithm of one minus absorptance as measured on a uniform sample.

**Fourier-transform infrared spectroscopy (FTIR):** Measurement technique whereby spectra are collected based on measurements of the temporal coherence of a radiative source (infrared radiation, in this case), using time-domain measurements of the electromagnetic radiation or other type of radiation.

**Gap (bandgap energy, $E_g$)**: The energy difference between the bottom of the conduction band and the top of the valence band in a semiconductor or an insulator.

**Interlaboratory Study:** Exercises performed by a number of independent laboratories to assess their performance and ability to deliver accurate results.

**Internal transmittance:** refers to energy loss by absorption within a particle (sample specimen), whereas the total transmittance is that due to absorption, reflection, scatter, etc.

**IR-FPA:** Infrared-Focal Plane array refers to an instrumental device used to measure infrared spectra whose detector is a combination of many small detectors (pixels) distributed (generally) in a square. This allows to capture simultaneously IR spectra in many spatial possitions.

**Mass spectrometry:** The study of systems by causing the formation of gaseous ions, with or without fragmentation, which are then characterized by their mass-to-charge ratios and relative abundances.

**Monochromatic (radiation):** Electromagnetic radiation consisting of a narrow range of wavelengths; consisting of photons with similar energy.

**Napierian absorbance, $A_e$:** The negative napierian logarithm of one minus absorptance as measured on a uniform sample: $A = -\log_n(1 - \alpha)$.

**Noise (instrumental):** The random fluctuations occurring in a signal that are inherent in the combination of instrument and method.

**Nyquist theorem:** The Nyquist theorem states that an analog signal can be digitized without aliasing error if and only if the sampling rate is greater than or equal to twice the highest frequency component in a given signal.

**Phase:** In the context of radiation, the phase refers to the relative position of a point in the wave. When considering two or more photons, they can be in-phase, if their electromagnetic waves are parallel, generating a "constructive interference", or out-of-phase, generating a "destructive interference".

**Photodegradation:** degradation by means of radiant energy (usually solar light).

**Polychromatic (radiation):** Electromagnetic radiation consisting of multiple wavelengths, consisting of photons with multiple different energies.

**Power** (which should be termed radiant flux or radiant power), is the radiant energy emitted, reflected, transmitted, or received per unit time, in watts) (it is confused often with brightness).

**Pyrolysis-gas chromatography:** A version of reaction chromatography in which a sample is thermally decomposed to simpler fragments before entering the column.

**Raman spectroscopy:** Spectroscopic technique based on the inelastic scattering of reflected radiation, usually ultraviolet, IR, visible or X-ray.

**Reflectance, $\boldsymbol{\rho}$**: Fraction of incident radiation reflected by a surface or discontinuity, $\rho(\lambda) = P_\lambda^{refl}/P_\lambda^0$, where $P_\lambda^0$ and $P_\lambda^{refl}$ are the incident and reflected spectral radiant power, respectively.

**Semiconductor laser:** A source of ultraviolet, visible, or infrared radiation which produces light amplification by stimulated emission of radiation from which the acronym is derived. The light emitted is coherent except for superradiance emission. The essential elements of a laser are: (i) an active medium; (ii) a pumping process to make a population inversion; and (iii) a suitable geometry of optical feedback elements. The active medium consists of a host material (gas, liquid or solid) containing an active species.

**Spectral intensity**, when brilliance (radiant intensity) is given per wavelength, watt per steradian per metre units.

**Spectral power density** (or spectral power distribution, power per unit area per unit wavelength), related to spectral intensity.

**Spectral power distribution** (related to spectral power density) can refer to the concentration, as a function of wavelength, of any radiometric or photometric quantity like radiant energy, radiant flux, radiant intensity, etc.

**Transmittance, T, τ:** The ratio of the transmitted radiant power ($P_\lambda$) to that incident on the sample ($P_\lambda^0$):

$$T = P_\lambda/P_\lambda^0$$

**Valence band:** Highest energy continuum of energy levels in a solid that is fully occupied by electrons at 0 Kelvin.

# 2.- SUPPLEMENTARY MATERIAL SECTION 2 OF THE PAPER

In section 2.2. of the main text it is explained that QCL-LDIR instruments perform reflectance-absorbance measurements. The analytical technique that uses the penetration of the light and its final emergence as a reflected beam receives several names, like IR reflection-absorption spectroscopy (IRRAS), reflection-absorption IR spectroscopy (RAIRS), IR reflection-absorption spectroscopy (IRAS), reflection-absorption spectroscopy (RAS) or just transflectance, as it was defined first time. The latter term was coined to epitomize what essentially happens during the measurement: the light beam passes through the sample twice, before and after its reflection at the totally reflective surface (in practical terms this will be a silver, aluminium or gold-coated reflective slide or filter), see Figure 3a of the main text, and in principle, the resulting spectrum can look like a transmittance one.

Following that discussion it is comented that despite the high intensity radiation emitted by QCL sources, it was reported that some detectors may reduce this advantage. In particular, microbolometers suffer from what it was called coherence artefacts, by which the QCL-based spectra look different from the FTIR ones due to speckles, fringes and interferences. As a result, it is difficult to know the part of the information that results from interferences and that resulting from absorbance. This effect is mostly associated to the use of microbolometers and patented algorithms allow to address this problem nowadays. When single point MCT detectors are used this problem is avoided.

Finally, also in Section 2.2 it is explained that reflectance measurements contain specular and diffuse components. The Agilent's LDIR instrumentation mitigates this problem by a built-in device that separates the specular and the diffuse reflectances by performing polarization measurements at a point of interest. To summarize, recall from section 2.1. that the laser beam of a QCL source is polarized (although additional linear polarization filters can be used afterthe laser beam is produced). Polarized light that undergoes specular reflection remains polarized whereas diffuse light gets depolarized. In this way the diffusely reflected light can be selectively measured with the aid of a linear polarization filter. Additional detailed information on the instrumental setup cannot be disclosed due to patent protection. Hence, the LDIR uses a set of orthogonal polarizers to measure the intensity of light proceeding from the measurement spot of a particle at different relative angles and using mathematical algorithms the specular and diffuse components of the light can be calculated automatically.

# 3.- SUPPLEMENTARY MATERIAL FIGURES

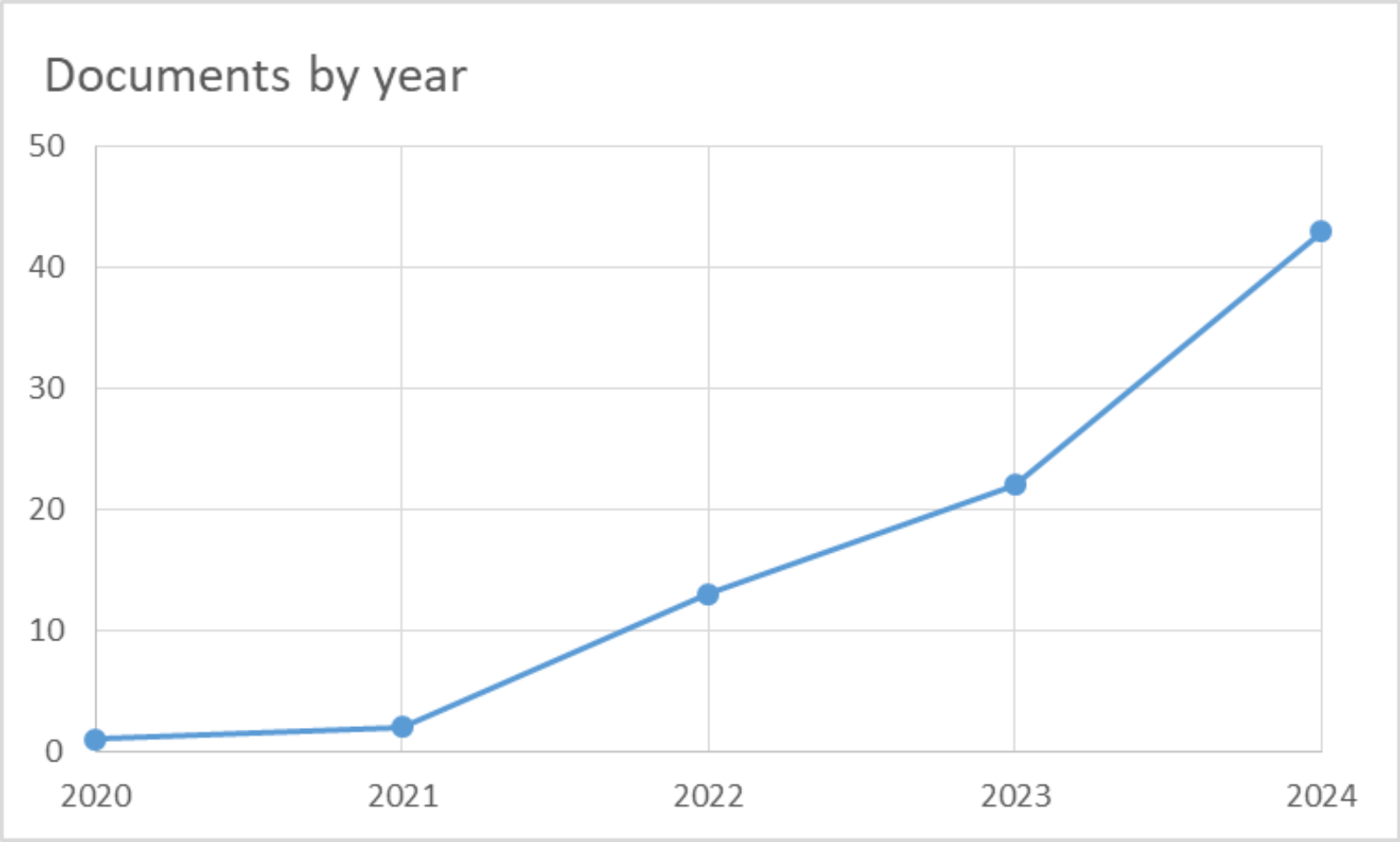

**Figure SM1:** Number of publications that use LDIR systems to perform microplastic-related studies as a function of the publication year. Source: Scopus, 3/February/2025. Search words: "microplastic" AND "LDIR".

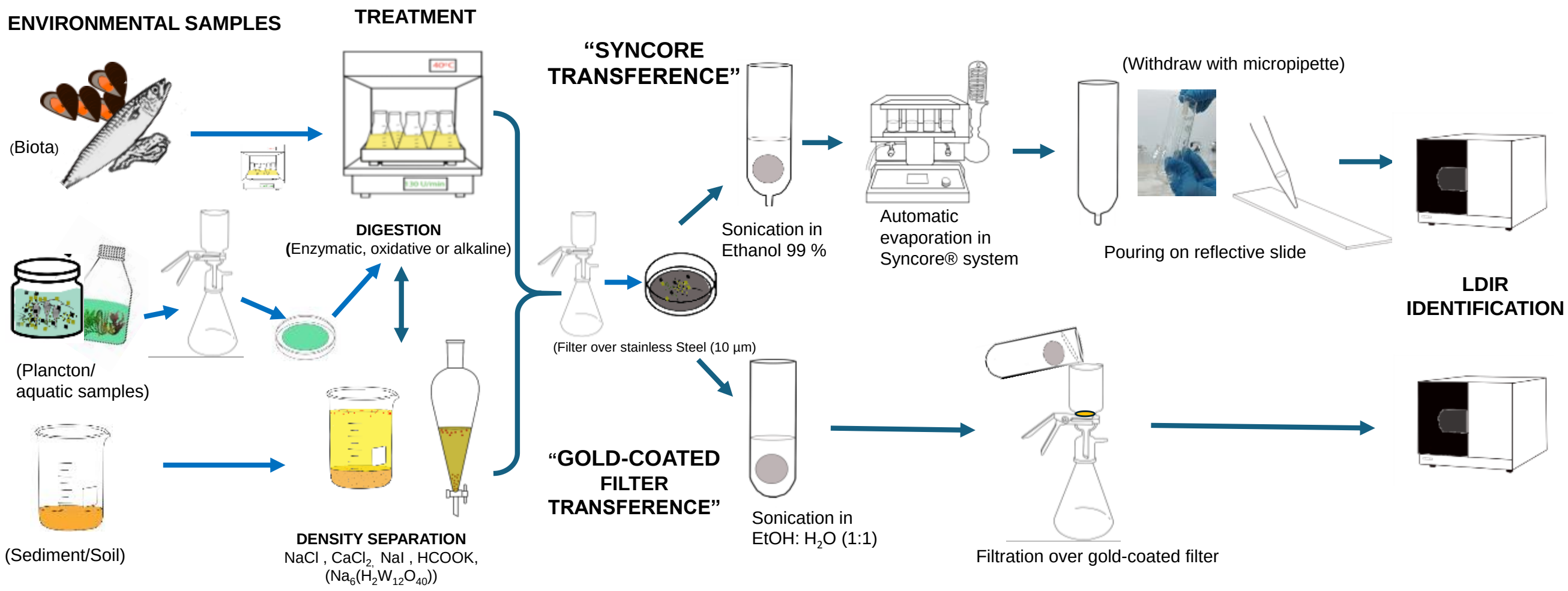

**Figure SM2(a):** General schemes for the sample treatment procedures followed in previous works undergone by the authors, referred to in the main text (see references from Lopez-Rosales et al.).

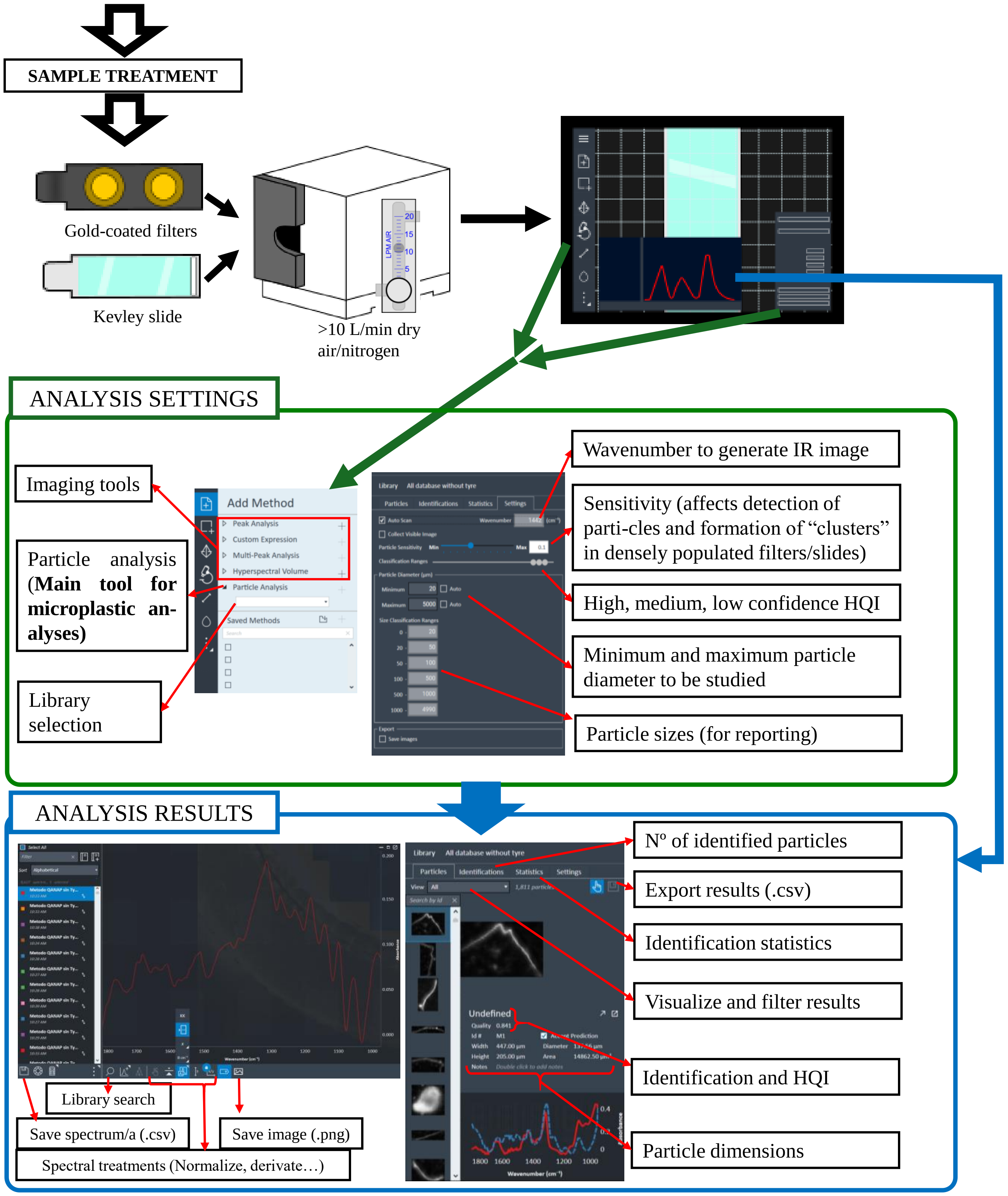

**Figure SM2(b):** General workflow to set the most relevant instrumental parameters for routine operation with an LDIR system.

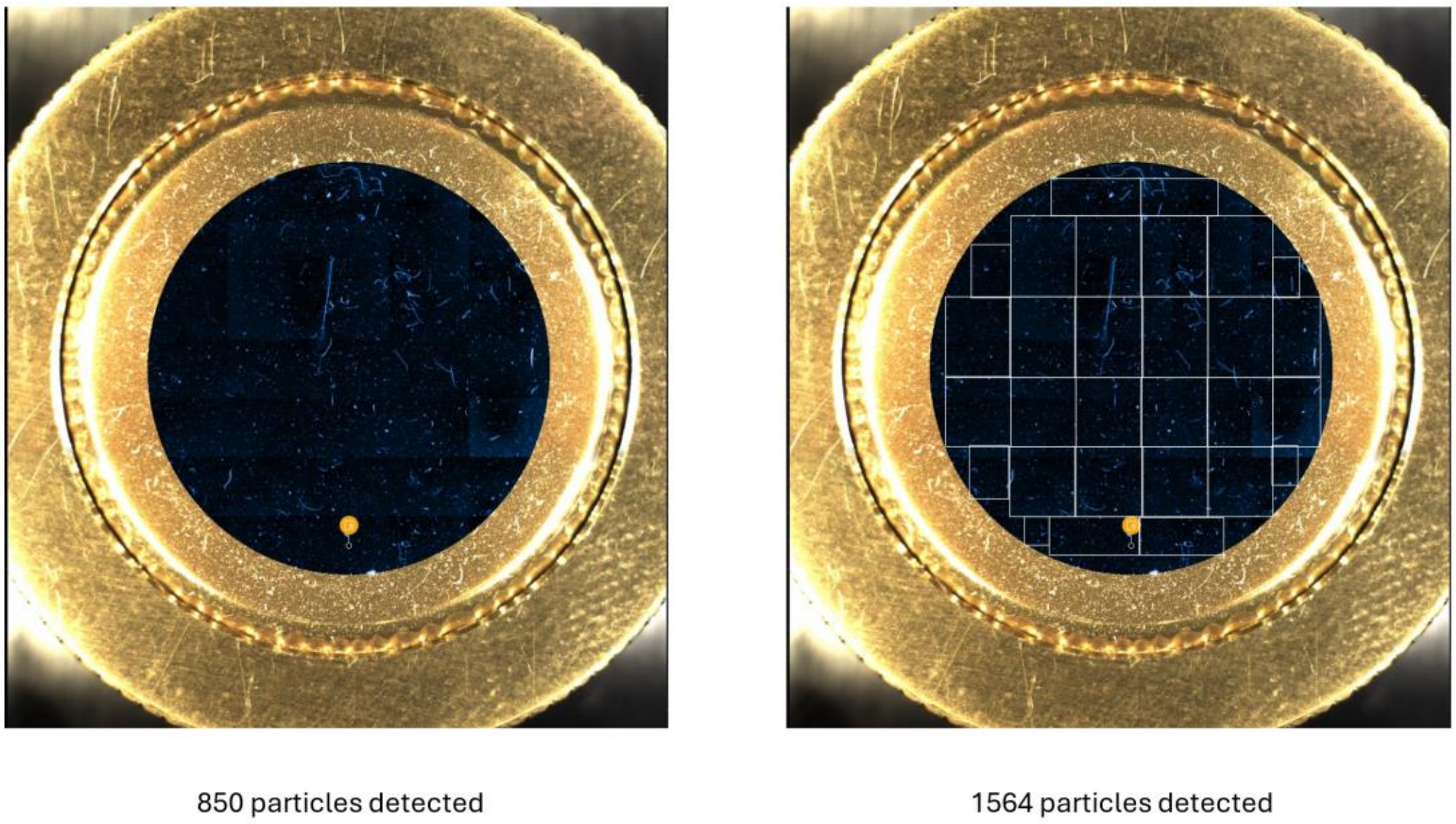

**Figure SM3**. Setting small areas for subsequent measurement increases particle detection.

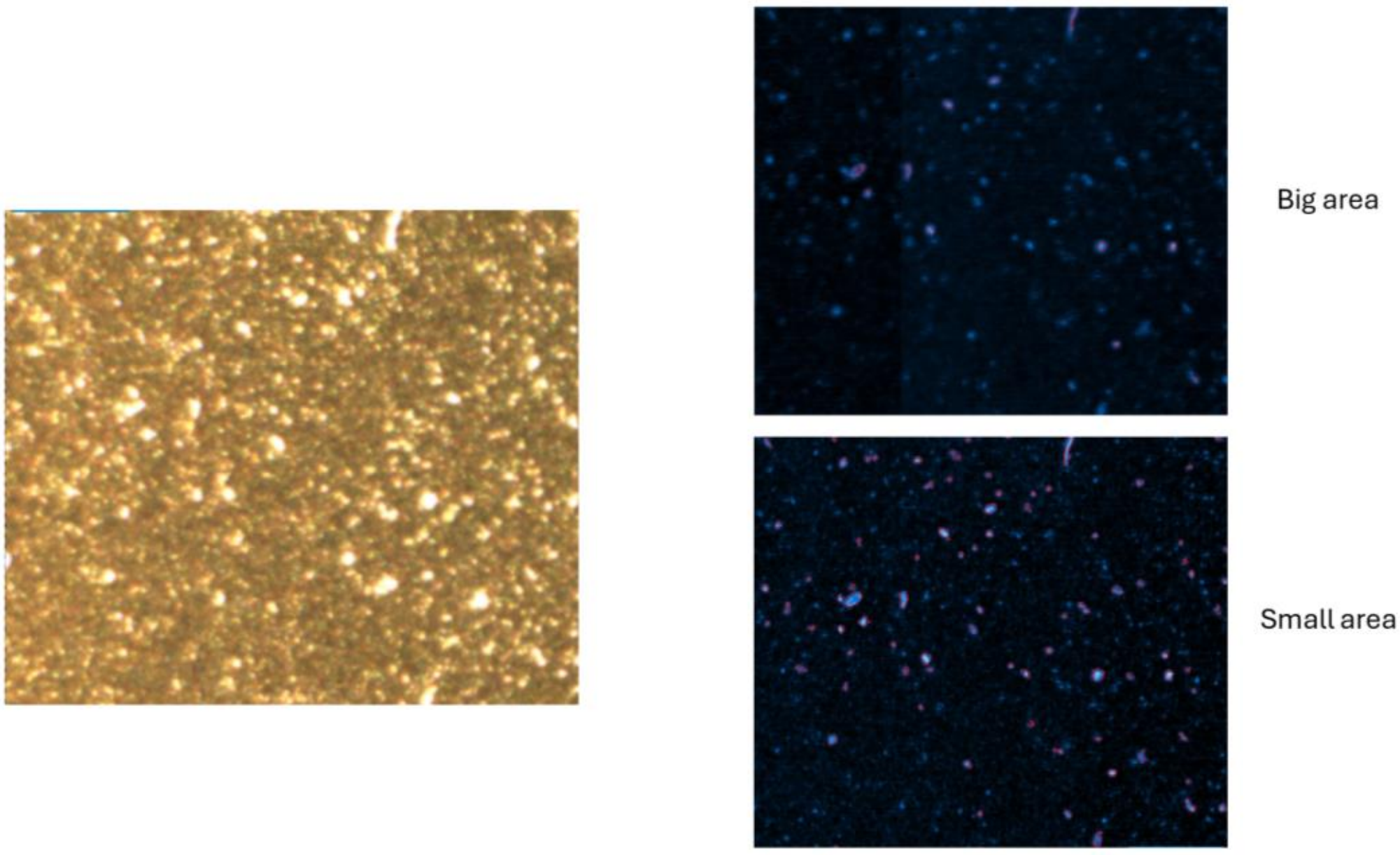

**Figure SM4**. Setting small areas allows more particles to be detected. The particles detected in the upper right side are marked in red in the bottom figure.